\documentclass[journal]{IEEEtran}
\usepackage{cite}
\usepackage{amsmath,amssymb,amsfonts}
\usepackage{epsfig}
\usepackage{graphics}
\usepackage{textcomp}
\def\BibTeX{{\rm B\kern-.05em{\sc i\kern-.025em b}\kern-.08em
    T\kern-.1667em\lower.7ex\hbox{E}\kern-.125emX}}

\usepackage[table,xcdraw]{xcolor}
\usepackage{algorithm, algpseudocode}
\newtheorem{lemma}{Lemma}

\newcommand{\argmin}[1]{\underset{#1}{\operatorname{arg}\,\operatorname{min}}\;}

\usepackage{url}

\ifCLASSINFOpdf
\else
\fi
\begin{document}
%
%
%
%

\title{Non-myopic GOSPA-driven Gaussian Bernoulli Sensor Management}

\author{George Jones, \'Angel F. Garc\'ia-Fern\'andez, Christian Blackman\\
\thanks{G. Jones, A. F. Garc\'ia-Fern\'andez and C. Blackman are with the Department of Electrical Engineering, and Electronics, University of Liverpool, Liverpool
L69 3GJ, U.K (emails: \{g.jones6, angel.garcia-fernandez, c.blackman\}@liverpool.ac.uk). A. F. Garc\'ia-Fern\'andez is also with the ARIES Research Centre, Universidad Antonio de Nebrija, Madrid, Spain.
This work was supported by the EPSRC Centre for Doctoral Training in Distributed Algorithms EP/S023445/1 and Roke Manor Research Limited.}}

\maketitle

\begin{abstract}
In this paper, we propose an algorithm for non-myopic sensor management for Bernoulli filtering, i.e., when there may be at most one target present in the scene. The algorithm is based on selecting the action that solves a Bellman-type minimisation problem, whose cost function is the mean square generalised optimal sub-pattern assignment (GOSPA) error, over a future time window. We also propose an implementation of the sensor management algorithm based on an upper bound of the mean square GOSPA error and a Gaussian single-target posterior. Finally, we develop a Monte Carlo tree search algorithm to find an approximate optimal action within a given computational budget. The benefits of the proposed approach are demonstrated via simulations.
\end{abstract}


\section{Introduction}
\IEEEPARstart{C}{onducting} surveillance of a specified area can be completed by utilising autonomous vehicles which have a limited Field of View (FOV). This FOV is typically significantly smaller than the desired surveillance region and therefore the vehicle has to be agile in a way which allows it to alter the positioning of the FOV, allowing it to view adjacent areas. This can be visualised as an autonomous ground vehicle traversing the ground in search of aerial targets \cite{Cormack2016}. Other work was completed in \cite{Gehly2018} looking at search-detect-track sensor management for geosynchronous space objects. The discipline that plans which actions the sensor/vehicle should take at each time-step to achieve an objective of interest, such as the tracking of these aerial targets, is referred to as sensor management \cite{Krishnamurthy_book16, Li2022}.

In this paper, we deal with sensor management for Bernoulli filtering \cite{Ristic2013}, in which, at any given time-step, a maximum of one target can be present. In a Bernoulli filter, a target is born, then moves with a  certain dynamic model, and then disappears from the surveillance area. Only once this target has disappeared, a new target may appear \cite{Hero_book08}. This target is observed through noisy measurements, which may contain clutter as well as target detections. \cite{Ristic2013}. In Bernoulli filtering, the posterior probability of target existence and the spatial density are propagated through the filtering recursion \cite{Ristic2013, Ristic_book04}. 

This paper deals with the problem of finding the optimal sequence of actions for an agile sensor platform to keep track of the current target until it disappears, and then search for a potential new target. In a model-based setting, this type of problem is usually posed as a Partially Observable Markov Decision Process (POMDP) which is a framework that allows for planning  when the system state is observed with uncertainty \cite{Hero_book08, Thrun_book05}. Another type of framework to solve sensor management problems is to use reinforcement learning (usually in combination with neural networks). In \cite{Oakes2022}, a double deep Q network is used to conduct sensor management on a ground based telescope for space situational awareness.  Reinforcement learning methods have also been used to approach multiple target tracking in \cite{Gorji2018}, \cite{Ren2018} and sensor management for single target tracking in \cite{Hoffmann2020}. These methods can achieve very high performance, but typically require long training times and may lack interpretability.

In this paper, we focus on POMDPs which have an interpretable methodology whose aim is to minimise a cost function (or maximise a reward function). The cost function can consider a single time-step ahead of the current time-step. This approach is referred to as myopic planning. Intuitively, there are limiting cases in which myopic planning cannot find a desirable solution, such as having to navigate around an obstacle that requires a multi-time step planning approach (non-myopic planning). We proceed to review the literature to address the sensor management problem.

A popular cost function for single-target tracking is the posterior Cram\'{e}r-Rao lower bound (PCRLB) \cite{Tichavsky1998}. The PCRLB has been applied to single-target tracking problems dealing with cluttered environments in \cite{Hernandez2006} and multiple-targets for multisensor array management in \cite{Tharmarasa2007}. It has also been used to develop a cognitive radar framework for target detection and tracking in \cite{Bell15} and used for extended target tracking in \cite{Tang2019}. The PCRLB is a bound on the mean square error. The PCRLB is used instead of the mean square error due to its good performance and the fact that it is computationally efficient to calculate. As the PCRLB is a bound on the mean square error, using it as a cost function must incorporate external criteria to be able to carry out tracking of an unknown and varying number of targets. 

Another method is to use a cost function based on information theory. Information-theoretic methods aim to
select the action that maximises the expected information gain between the predicted and posterior density. Their objective is to take the action that gives the most information about the variables of interest. The information gain is measured by an information theoretic divergence, such as the Kullback-Leibler (KL) divergence or the R\'{e}nyi divergence \cite{Kreucher07, Aoki11, Ristic11b, Beard17b, LeGrand2023, Saucan2021}. Whilst these approaches can lead to desirable results, it is not explicitly clear what the resulting policy is aiming for in terms of more practical aspects of multiple target-tracking.

The non-myopic case of sensor management considers the longer term impact of the actions that are chosen now \cite{Kreucher2004}. The ability to plan further into the future, whilst offering many benefits, also carries some drawbacks. Building non-myopic sensor management algorithms for POMDP's can be challenging as the state is not fully observable, meaning both the state of the target and state of the sensor have to be updated without knowing whether or not an observation will be received. A proposed solution using Monte Carlo Tree Search for managing large-scale Partially Observable Markov Decision Processes is detailed in \cite{Silver2010}. As well as this, the problem suffers from combinatorial explosion, meaning the problem quickly becomes intractable, even with moderate planning horizons. An efficient way of searching the action space has been proposed in \cite{Chhetri2004} and a distributed approach to non-myopic planning for multiple targets in \cite{Park2019}. Non-myopic planning has also been used for the control of a single pan-tilt-zoom camera using information theoretic drivers in \cite{Salvagnini2015}. A related problem to non-myopic sensor management is the control of an unknown number of interceptors to rendezvous with a given target at a given time \cite{Clark2022}.

In this paper, we address the sensor management problem by considering a cost function based on a metric for sets of targets. A sensor management algorithm based on the optimal sub-pattern assignment metric was proposed in \cite{Gostar2017}. In this work, we use the generalised optimal sub-pattern assignment (GOSPA) metric \cite{Rahmathullah2017}.  
The GOSPA metric penalises for the localisation error for properly detected  targets, the missed targets error and the false targets error \cite{GarciaFernandez2021}, which are concepts of interest in traditional multi-target tracking performance evaluation \cite{Fridling91}. The GOSPA metric combines these three quantities in a mathematically principled manner providing a cost function that can be used to quantitatively measure performance. The GOSPA localisation error resembles the (unnormalised) multiple object tracking precision score in \cite{Bernardin2008}. The GOSPA metric has been used in track-before-detect sensor management for Bernoulli filtering in \cite{Ubeda_thesis18} and in a non-myopic sample-based approximation for Bernoulli filtering in \cite{Hernandez2023}. An analysis of the favourable properties of the GOSPA metric compared to other metrics for sets of targets in the context of sensor management is provided in \cite{GarciaFernandez2021}.

In this paper, we extend the work from \cite{Jones2023} to include non-myopic planning, using GOSPA as a driver of performance in a Gaussian Bernoulli setting. Specifically, we propose an upper bound on the mean square GOSPA (MSGOSPA) as a cost function to assign a cost to each available action. We use this bound as it is closed-form for linear-Gaussian systems and computationally efficient to calculate. Then, we develop a Monte Carlo Tree Search (MCTS) algorithm \cite{Browne2012} that enables us to solve the non-myopic planning problem in a computationally efficient manner. This algorithm is benchmarked against an information theoretic approach - using the KL divergence. Preliminary results were provided in the conference version of this paper in \cite{Jones2023}.\\

In summary, the contributions of this paper are:

\begin{enumerate}
    \item A sensor management framework for Bernoulli filtering based on GOSPA, suitable for myopic and non-myopic planning.
    \item The development of a closed form cost function for sensor management based on an upper bound of the MSGOSPA error for Gaussian single-target distributions.
    \item A computationally efficient implementation of the planning algorithm based on an MCTS method
    \item An approximation of the expected probability of detection based on importance sampling for the case in which sensors have a circular FOV with a constant probability of detection.
\end{enumerate}

The rest of the paper is organised as follows. Section \ref{sec: problem formulation}
defines the problem of sensor management, Section \ref{sec: non-myopic} details the non-myopic Gaussian Bernoulli sensor management problem, Section \ref{sec: MCTS} explains the implementation of the MCTS and how it is applied to this problem, Section \ref{sec: numerical experiments} provides the parameters and results of the conducted simulations and Section \ref{sec: conclusion} provides concluding statements and planned avenues for future developments in this work.
\section{Problem Formulation \& Background}\label{sec: problem formulation}
In this section, we provide the sensor management problem formulation and the required background. In particular, in Section \ref{sec: Bernoulli and measurement model}, we introduce the dynamic and measurement models for Bernoulli filtering. In Section \ref{sec: Bernoulli filtering}, we outline the Bernoulli filtering recursion. In Section \ref{sec: GOSPA metric}, we review the GOSPA metric. In Section \ref{sec: non-myopic planning and time-discounted predicted MSGOSPA error}, we discuss non-myopic planning using the time-discounted predicted MSGOSPA error.
\subsection{Bernoulli Dynamic \& Measurement Model}\label{sec: Bernoulli and measurement model}
In a Bernoulli model, the target may or may not be present in the surveillance area at a given time-step $k$. That is, the multi-target state at time-step $k$ is a set $X_k$ that can either be empty ($X_k = \emptyset$) or a singleton ($X_k = \{x_k\}$) where $x_k\in \mathbb{R}^{n_{x}}$ is the single-target state \cite{Mahler_book14}.

The dynamics of a Bernoulli Markov process are characterised by the multi-target transition density $\phi_{k|k-1}(X_k|X_{k-1})$. For $X_{k-1} = \emptyset$, the transition density is
\begin{equation}\label{eqn: transitional PDF emptyset}
    \phi_{k|k-1}(X_k|\emptyset) = 
    \begin{cases}
      1 - p^B & X_k = \emptyset\\
      p^{B} \cdot b_{k|k-1}(x_k) & X_k = \{x_k\} \\
      0 & |X_k| \geq 2\\
    \end{cases}
\end{equation}
where $p^B$ is the probability of birth, $b_{k|k-1}(x_k)$ is the single-target birth density at time-step $k$ and $|X_k|$ is the cardinality of the state set.  

For $X_{k-1} = \{x_{k-1}\}$, the transition density is
\begin{multline}\label{eqn: transitional PDF full set}
    \phi_{k|k-1}(X_k|\{x_{k-1}\})\\=
    \begin{cases}
      1 - p^S & X_k = \emptyset\\
      p^S \cdot
      \pi_{k|k-1}(x_k|x_{k-1}) & X_k = \{x_k\} \\
      0 & |X_k| \geq 2\\
    \end{cases}
\end{multline}
where $p^S$ is the probability of survival and $\pi_{k|k-1}(x_k|x_{k-1})$ is the single-target transition density.
The birth density contains information on where a new target could appear, and the single-target transition density contains information on how the present target moves.

The sensor has the ability to select a sensing mode $a_{k}\in\mathbb{A}$ to sense the environment at each time-step $k$, where $\mathbb{A}$ is a finite set with all of the available sensing modes. At each time-step, a target  $x\in X_{k}$ has the possibility of being detected with a probability of detection $p_{a_{k}}^{D}\left(\cdot\right)$. A target generated measurement $z$ is then produced with a linear Gaussian density $l(z|x)=\mathcal{N}\left(z;H_{a_{k}}x+b_{a_{k}},R_{a_{k}}\right)$, in which $H_{a_k}$ is the observation matrix, $R_{a_k}$ is the observation noise matrix and $b_{a_{k}}$ is a bias term. The notation $\mathcal{N}\left(z;\hat{z},S\right)$ denotes a Gaussian density, evaluated at $z$, and parameterised by $\hat{z}$ and $S$ - a mean and a covariance matrix respectivley. Clutter is generated by a Poisson point process with clutter rate (intensity) $\lambda_{c}(\cdot)$.
The set of measurements at time-step $k$, which can contain both clutter and a target generated measurement is denoted $Z_k$.
\subsection{Bernoulli Filtering}\label{sec: Bernoulli filtering}
%
%
In Bernoulli filtering, both the predicted and posterior densities are Bernoulli densities of the form
\begin{equation}\label{eqn: bernoulli density}
    f_{k|k'}(X_k) =
    \begin{cases}
      r_{k|k'}p_{k|k'}(x_{k-1}) & X_k = \{x_k\}\\
      1 - r_{k|k'} & X_k = \emptyset\\
      0 & |X_k|  \geq  2
    \end{cases}
\end{equation}
where $k' \in \{k-1, k\}$ with $k'=k-1$ for the predicted density and $k'=k$ for the posterior density, $p_{k|k'}(\cdot)$ is the single-target density and $r_{k|k'}$ is the probability of existence.

The Bernoulli filtering recursion propagates the probability of existence and the single-target density. The prediction equations are given in \cite{Ristic2013}, see Eqs. (28) and (29). The update equations are given by Eqs. (56), (57) and (59) in \cite{Ristic2013} and have not been repeated here for brevity.
\subsection{The GOSPA metric}\label{sec: GOSPA metric}
Given two sets of targets $X$ and $Y$, an assignment set $\gamma$ between them has the following properties $\gamma \subseteq \{ 1,...,|X|\} \times \{1,...,|Y|\}, (i,j), (i,j') \in \gamma \Longrightarrow j=j' \quad \text{and} \quad (i,j),(i',j) \in \gamma \Longrightarrow i = i'$. The last two properties ensure that each target is assigned at most once. Given the parameters $\alpha = 2$, maximum localisation error $c > 0$, $p > 0$ and a base metric in the single-target space $d(\cdot, \cdot)$, the GOSPA metric is \cite{Rahmathullah2017}
\begin{multline}
    d_p^{(c,2)} (X,Y)\\
    = \underset{\gamma\in\Gamma}{\text{min}}\left( \sum\limits_{(i,j) \in \gamma} d^p(x_i,y_j) + \frac{c^p}{2}(|X| + |Y| - 2|\gamma|)\right)^{1/p}
\end{multline}
where $d^p(\cdot)$ is the localisation error to the $p$-th power. In a target-tracking scenario, the sets $X$ and $Y$ represent the ground truth and estimated set, respectively. Then, $|X| - |\gamma|$ is the number of missed targets and $|Y| - |\gamma|$ is the number of false targets.

For notational simplicity, in the rest of the paper, we drop the superindices and subindices in the GOSPA metric and just denote it as $d(X,Y)$. We also assume that $p=2$ and that the base metric is the Euclidean metric.
\subsection{Non-myopic Planning using the time-discounted predicted MSGOSPA Error}\label{sec: non-myopic planning and time-discounted predicted MSGOSPA error}
In non-myopic sensor management, we optimise over a policy that minimises the predicted cost over multiple future time-steps \cite[Chapter 7]{Krishnamurthy_book16}. All available information up to time-step $k' \geq k$ for sensor management at time-step $k$ is denoted as
\begin{equation}
\mathcal{I}_{k'} = 
    \begin{cases}
        (f_{k|k-1}(\cdot)) & k'= k \\
        (f_{k'|k'-1}(\cdot), a_{k:k'-1}, Z_{k:k'-1}) & k'\geq k
    \end{cases}
\end{equation}
where $a_{k:k'-1}=(a_{k},...,a_{k'-1})$ and $Z_{k:k'}=(Z_{k},...,Z_{k'-1})$ denote the sequences of actions and measurement sets from time-step $k$ to $k'-1$, respectively. The agent then makes decisions, considering the dynamic and measurement models, according to a deterministic policy $\mu_{k'}(\cdot)$ that maps the available information to the next action such that
\begin{equation}
    a_{k'} = \mu_{k'}(\mathcal{I}_{k'})
\end{equation}
Considering a planning horizon up to time-step $K$, with a length of $K-k+1$, the sequence of policies up to the planning horizon is denoted by $\mu_{k:K}(\cdot) = (\mu_{k}(\cdot), \ldots, \mu_{K}(\cdot))$. Including a decay factor $\lambda \in (0,1]$, for GOSPA-driven sensor management, the policy is chosen to minimise \cite[Eq. (7.6)]{Krishnamurthy_book16} 
\begin{multline}\label{eqn: J not nested}
    J_{\mu_{k:K}(\cdot)} = \\
    E_{\mu_{k:K}(\cdot)}\left[ \sum^K_{k'=k} \lambda^{k'-k}d^2\left(X_{k'}, \hat{X}_{k'}(a_{k:k'}, Z_{k:k'})\right)\right]
\end{multline}
where the expectation is taken with respect to the joint probability density of $(X_k, \ldots, X_K, Z_k,\ldots,Z_K)$ under the policy $\mu_{k:K}(\cdot)$, and $\hat{X}_{k'}(a_{k:k'},Z_{k:k'})$ is the optimal MSGOSPA estimator at time-step $k'$.\\

The optimal policy is then
\begin{equation}\label{eqn: the optimal policy}
    \mu_{k:K}^*(\cdot) = \argmin{\mu_{k:K}(\cdot)} J_{\mu_{k:K}(\cdot)}
\end{equation}
and the minimum value of the cost is
\begin{equation}
    J_{\mu^*_{k:K}(\cdot)} = \min_{\mu_{k:K}(\cdot)} J_{\mu_{k:K}(\cdot)}
\end{equation}
The minimum value of the cost can also be written more explicitly as the nested minimisations in \eqref{eqn: J nested}-\eqref{eqn: J nested part 2},
\begin{table*}
\centering
\begin{minipage}{0.9\textwidth}
\hrule
\begin{align}\label{eqn: J nested}
J_{\mu_{k:K}^{*}\left(\cdot\right)} & =\underset{a_{k}}{\min}\,\mathrm{E}_{Z_{k};a_{k}}\left[d^{2}\left(X_{k},\hat{X}_{k}\left(a_{k},Z_{k}\right)\right)\right.\nonumber \\
 & \quad+\lambda\underset{a_{k+1}}{\min}\,\mathrm{E}_{Z_{k+1}|Z_{k};a_{k:k+1}}\left[d^{2}\left(X_{k+1},\hat{X}_{k+1}\left(a_{k:k+1},Z_{k:k+1}\right)\right)\right.\nonumber \\
 & \left.\left.\quad+...+\lambda^{K-k}\underset{a_{K}}{\min}\,\mathrm{E}_{Z_{K}|Z_{k:K-1};a_{k:K}}\left[d^{2}\left(X_{K},\hat{X}_{K}\left(a_{k:K},Z_{k:K}\right)\right)\right]...\right]\right].\\
 & =\underset{a_{k}}{\min}\,\int f_{k|k-1}^{m}\left(Z_{k};a_{k}\right)\left[\int d^{2}\left(X_{k},\hat{X}_{k}\left(a_{k},Z_{k}\right)\right)f_{k|k}\left(X_{k}\left|Z_{k};a_{k}\right.\right)\delta X_{k}\right.\nonumber \\
 & \quad+\lambda\underset{a_{k+1}}{\min}\int f_{k+1|k}^{m}\left(Z_{k+1}|Z_{k};a_{k:k+1}\right)\left[\int d^{2}\left(X_{k+1},\hat{X}_{k+1}\left(a_{k:k+1},Z_{k:k+1}\right)\right)f_{k+1|k+1}\left(X_{k+1}\left|Z_{k:k+1};a_{k:k+1}\right.\right)\delta X_{k}\right.\nonumber \\
 & \quad+...+\lambda^{K-k}\underset{a_{K}}{\min}\,\int f_{K|K-1}^{m}\left(Z_{K}|Z_{k:K-1};a_{k:K}\right) \nonumber \\
 & \left[\int d^{2}\left(X_{K},\hat{X}_{K}\left(a_{k:K},Z_{k:K}\right)\right)f_{K|K}\left(X_{K}\left|Z_{k:K};a_{k:K}\right.\right)\delta X_{K}\right]\delta Z_{K}...\bigg]\delta Z_{k+1}\bigg]\delta Z_{k}. \label{eqn: J nested part 2}
\end{align}
\medskip
\hrule
\end{minipage}
\end{table*}
where each minimisation provides the optimal action at a time-step for each possible sequence of past measurements and actions. The integrals in \eqref{eqn: J nested part 2} correspond to set integrals in \cite{Mahler_book14}, which are reviewed in Appendix \ref{appendix A}. The terms in \eqref{eqn: J nested part 2} are defined as $f^m_{k|k-1}(\cdot ; a_k)$ being the predicted density of the measurements for action $a_k$ and $f_{k|k}(\cdot|Z_k,a_k)$ is the posterior density given action $a_k$ and measurement set $Z_k$. These two densities are explained in more detail in Sections \ref{sec: predicted measurement density} and \ref{sec: updated Bernoulli density}, respectively. 
The nested expectations can then be written in a Bellman-type equation using the value function \cite{Krishnamurthy_book16}. That is, the value function at the final time-step $K$ is
\begin{multline}
   V_K(Z_{k:K-1} , a_{k:K-1}) = \\
    \min_{a_K}E_{Z_K|Z_{k:K-1},a_{k:K-1}}\left[d^2\left(X_K, \hat{X}_K(a_{k:K}, Z_{k:K})\right)\right]
\end{multline}
where it should be noted that the value function depends on previous measurements and actions.

For $k' \in \{k, k+1, ... , K-1 \}$, the value function can be computed recursively backwards via
\begin{multline}
    V_{k'}(Z_{k:k'-1}; a_{k:k'-1} ) = \\
    \min_{a_{k'}}E_{Z_{k'}|Z_{k:k'-1};a_{k:k'}} \bigg[ d^2\left( X_{k'}, \hat{X}_{k'}(a_{k:k'},Z_{k:k'}) \right) \\ 
    + \lambda^{k'+1-k}V_{k'+1}(Z_{k:k'};a_{k:k'}) \bigg]
\end{multline}
A special case of the non-myopic planning can be considered in which there is no multi-step lookahead, either setting $K=k$ or $\lambda=0$ in \eqref{eqn: J not nested}. The actions decided by the sensor are informed only by the predictions one time step ahead of the current time step. This is known as myopic planning and results in this optimal action at time step $k$ \cite{Hernandez_inbook13}.
\begin{multline}\label{eqn: optimal myopic action}
    {a}_{k}^* = \underset{a_k}{\text{argmin}}\quad \mathrm{E} \left[ d^2 \left( X_k , \hat{X}_k (a_k, Z_k) \right) ; a_k\right]\\
    = \underset{a_k}{\text{argmin}} \int \bigg[\int d^2\left(X_k,\hat{X}_k(a_k, Z_k)\right) \\
    f_{k|k}(X_k|Z_k;a_k)\delta X_k\bigg]f_{k|k-1}^m(Z_k;a_k)\delta Z_k
\end{multline}
\section{Non-myopic Gaussian Bernoulli Sensor Management}\label{sec: non-myopic}
This section presents a sensor management algorithm for Bernoulli filtering based on the Gaussian distributions and the GOSPA metric. The assumptions of the sensor management algorithm are
\begin{enumerate}
    \item The clutter intensity $\lambda_c(\cdot)$ is zero. \label{assumption 1}
    \item We either detect zero measurements $Z_k = \emptyset$ or one measurement $Z_k = \{z_k\}$ at the predicted mean. \label{assumption 2}
    \item We only consider the component in the Gaussian mixture of the predicted single-target density with the highest associated weight. \label{assumption 3}
    \item The probability of detection is approximated as a constant given by its predicted value. \label{assumption 4}
    \item We use a computationally efficient upper bound for the resulting MSGOSPA error. \label{assumption 5}
\end{enumerate}
The rest of the section is organised as follows. Section \ref{sec: gaussian predictd density} presents the Gaussian predicted density, Section \ref{sec: predicted measurement density} presents the predicted measurement density, Section \ref{sec: updated Bernoulli density} presents the updated Bernoulli density for both cases where a measurement is received, and not received, Section \ref{sec: upper bound} presents the upper bound on the MSGOSPA error and Section \ref{sec: simplified bellamn eqn} presents the resulting Bellman equation.
\subsection{Gaussian Predicted Density}\label{sec: gaussian predictd density}
We first explain the form of the predicted density. Under Assumption 3), the predicted density is Bernoulli whose density is given by \eqref{eqn: bernoulli density} and where the single-target density is Gaussian where
\begin{equation}\label{eqn: predicted density}
    p_{k|k-1}(x_{k}) = \mathcal{N}(x_k;\bar{x}_{k|k-1}, P_{k|k-1})
\end{equation}
where $\bar{x}_{k|k-1}$ is the predicted mean and $P_{k|k-1}$ is the predicted covariance.
\subsection{Predicted Measurement Density} \label{sec: predicted measurement density}
Using Assumption 4), the probability of detection is approximated as
\begin{align}\label{eqn: expected probability of detection}
\overline{p}_{a_{k}}^{D}\big(\bar{x}_{k|k-1}, & P_{k|k-1} \big) \nonumber \\ 
& = E\left[{p}_{a_{k}}^{D}\left(x\right)\right] \nonumber \\ 
& = \int {p}_{a_{k}}^{D}\left(x\right)\mathcal{N}\left(x;\overline{x}_{k|k-1},P_{k|k-1}\right)dx
\end{align}
It can be noted here that the integral in \eqref{eqn: expected probability of detection} is over the single-target space $\mathbb{R}^{n_x}$.
Additionally, using Assumption 1), the predicted density of the measurement is Bernoulli with
\begin{multline}\label{eqn: predicted density of the measurement}
    f_{k|k-1}^m({Z}_k;a_k) \simeq\\
    \begin{cases}
      \overline{p}^D_{a_k}(\bar{x}_{k|k-1}, {P}_{k|k-1}) r_{k|k-1} \mathcal{N}(z_k;, \hat{z}_{a_k},S_{a_k}) & {Z}_k = \{z_k\}\\
      1 - r_{k|k-1} \overline{p}^D_{a_k}(\bar{x}_{k|k-1}, {P}_{k|k-1}) & {Z}_k = \emptyset 
    \end{cases}
\end{multline}
where the predicted measurement and its covariance matrix are
\begin{align}
    &\hat{z}_{a_k} = H_{a_k} \bar{x}_{k|k-1} + b_{a_k} \label{eqn: z_hat} \\
    & S_{a_k} = H_{a_k} P_{k|k-1} H_{a_k}^T + R_{a_k} \label{eqn: S_ak}
\end{align}
\subsection{Gaussian Posterior}
\label{sec: updated Bernoulli density}
Considering Assumption 2), we only need to consider two cases to compute the posterior: either there is a target-generated measurement $Z_{k} = \{\hat{z}_{a_k}\}$, or not $Z_{k} = \emptyset$. Using the Bernoulli update \cite[(6)-(8)]{Jones2023} for $\lambda_c(\cdot)\rightarrow 0$, we obtain the updated parameters.
%
%
\subsubsection{For ${Z}_k = \emptyset$}
The updated mean, covariance and probability of existence are
\begin{equation}\label{eqn: x^0}
    \bar{x}_{k|k,{a_k}}^0 = \bar{x}_{k|k-1} 
\end{equation}
\begin{equation}\label{eqn: P^0}
    P_{k|k,{a_k}}^0 = P_{k|k-1}
\end{equation}
\begin{equation}\label{eqn: r_k|k^0}
    r_{k|k,{a_k}}^0 = \frac{(1-\overline{p}^D_{a_k}(\bar{x}_{k|k-1}, {P}_{k|k-1}))r_{k|k-1}}{1 - r_{k|k-1} + (1 - \overline{p}^D_{a_k}(\bar{x}_{k|k-1}, {P}_{k|k-1}))r_{k|k-1}}
\end{equation}
%
%
\subsubsection{For ${Z}_k = \{\hat{z}_{a_k}\}$}
Under Assumption 1), the updated mean, covariance and probability of existence are \cite[App. A]{Jones2023}
\begin{equation}\label{eqn: x^1}
    \bar{x}_{k|k,{a_k}}^{1} = \bar{x}_{k|k-1} 
\end{equation}
\begin{equation}\label{eqn: P^1}
    P_{k|k,{a_k}}^{1} = P_{k|k-1} - P_{k|k-1} H_{a_k}^T (S_{a_k})^{-1} H_{a_k} P_{k|k-1}
\end{equation}
\begin{equation}\label{eqn: yes measurement prob existence}
    r_{k|k,{a_k}}^1 = 1
\end{equation}
The superscript $0$ in the updated Bernoulli parameters denote a misdetection hypothesis (no measurement received) and superscript $1$ denotes a measurement hypothesis (measurement received)
\subsection{Upper Bound on the MSGOSPA error}\label{sec: upper bound}
Calculating the  MSGOSPA error is intractable, and therefore we need to resort to approximations for its use in sensor management. This section provides a closed-form upper bound of the MSGOSPA error in linear Gaussian systems that is fast to calculate and suitable for sensor management.

Given an updated Bernoulli density with parameters \eqref{eqn: x^0}-\eqref{eqn: P^1}, let us consider the following estimator of the set of targets
\begin{equation} \label{eqn: estimator}
    \hat{X}\left(a_{k},Z_{k}\right)
    =
    \begin{cases}\left\{ \bar{x}_{k|k,a_{k}}^{|Z_{k}|}\right\} & r_{k|k,a_{k}}^{|Z_{k}|}\geq\Gamma_{d}\\
    \emptyset & r_{k|k,a_{k}}^{|Z_{k}|}<\Gamma_{d}
\end{cases}
\end{equation}
where $\Gamma_{d}$ is the detection threshold. That is, $\hat{X}\left(a_{k},Z_{k}\right)$ estimates a target located at the posterior mean if the updated probability of existence is greater than $\Gamma_{d}$ and an empty set if the probability of existence is lower. Then, the upper bound for the MSGOSPA error for this estimator is provided in Lemma 1 below. \vspace{1em}
\noindent\hrule \vspace{2pt}
\hrule

\vspace{3pt}
\begin{lemma}\label{lemma 1}
Let $(r_{k|k,a_k}^j, P^j_{k|k,a_k} )$ be the updated probability of existence and covariance matrix of the target for $|Z_k|=j, j\in \{0,1\}$. An upper bound on the MSGOSPA error for a given measurement set $Z_k$ is
\begin{multline}\label{eqn: upper bound on MSGOSPA error}
    \int d^{2}\left(X_{k},\hat{X}_{k}\left(a_{k},Z_{k}\right)\right)f_{k|k}\left(X_{k}\left|Z_{k};a_{k}\right.\right)\delta X_{k}\\
     \leq C(\Gamma_d,  r_{k|k,{a_k}}^{|{Z}_k|}, P_{k|k,{a_k}}^{|{Z}_k|})
\end{multline}
where
\begin{multline}\label{eqn: decomposed cost}
    C(\Gamma_d,  r_{k|k,{a_k}}^{|{Z}_k|}, P_{k|k,{a_k}}^{|{Z}_k|}) \\ =
    \begin{cases}
      \frac{c^2}{2}r_{k|k,{a_k}}^{|{Z}_k|} & r_{k|k,{a_k}}^{|{Z}_k|} \leq \Gamma_d\\
      \frac{c^2}{2}(1 - r_{k|k,{a_k}}^{|{Z}_k|}) & \\
      + r_{k|k,{a_k}}^{|{Z}_k|} \min \left(\text{tr}(P_{k|k,{a_k}}^{|{Z}_k|}),c^2\right) &  r_{k|k,{a_k}}^{|{Z}_k|} > \Gamma_d\\
    \end{cases}   
\end{multline}
A proof of this lemma is provided in Appendix \ref{Appendix B}.
\vspace{1em}
\noindent\hrule \vspace{2pt}
\hrule
\end{lemma}
%
%
\vspace{1em}
The upper bound given by Lemma 1 has two entries, depending on whether the updated probability of existence $r_{k|k,{a_k}}^{|{z}_k|}$ is greater than the detection threshold, $\Gamma_d$, or not. Interestingly, this depends only on whether there is a detection or not. It does not depend on either the posterior mean or the received measurement itself. The optimal detection threshold can also be obtained from Lemma 1 \cite[App. C]{Jones2023}
\begin{equation}
    \label{eqn: optimal detection threshold}
    \Gamma_d^* = \frac{1}{2 - \min \left(2 \frac{\text{tr}(P_{k|k,{a_k}}^{|{Z}_k|})}{c^2},1\right)}
\end{equation}
Substituting \eqref{eqn: optimal detection threshold} into \eqref{eqn: decomposed cost}, the MSGOSPA error upper bound for the optimal detection threshold becomes
\begin{multline}\label{eqn: final cost split}
    C(r_{k|k,{a_k}}^{|{Z}_k|}, P_{k|k,{a_k}}^{|{Z}_k|}) = \\
    \begin{cases}
      \frac{c^2}{2}r_{k|k,{a_k}}^{|{Z}_k|} & r_{k|k,{a_k}}^{|{Z}_k|} \leq \Gamma_d^*\\
      \frac{c^2}{2}(1 - r_{k|k,{a_k}}^{|{Z}_k|}) \\
      + r_{k|k,{a_k}}^{|{Z}_k|} \min \left(\text{tr}(P_{k|k,{a_k}}^{|{Z}_k|}),c^2\right) &  r_{k|k,{a_k}}^{|{Z}_k|} > \Gamma_d^*\\
    \end{cases}   
\end{multline}
\subsection{The Bellman Equation} \label{sec: simplified bellamn eqn}
In this section, we write the Bellman equation \eqref{eqn: J nested}-\eqref{eqn: J nested part 2} under the assumptions stated at the beginning of Section \ref{sec: non-myopic}. To do so and to simplify the notation in the nested integrals in \eqref{eqn: J nested part 2} when we work under these assumptions, we first introduce binary variable $o_k \in \{0,1\}$ to indicate when we are in a detection hypothesis $o_k = 1$, which implies $Z_k=\{\hat{z}_{a_k}\}$, or in a misdetection hypothesis $o_k = 0$, which implies $Z_k = \emptyset$. Then, $p(o_{k'}|a_{k:k'}, o_{k:k'-1})$ represents the probability of observing a measurement at time-step $k'$ given actions up to time-step $k'$ and past observations and is given by
\begin{multline}\label{eqn: observing a measurement density}
p(o_{k'}|a_{k:k'}, o_{k:k'-1}) = \\
    \begin{cases}
        \overline{p}^D_{a_k'}\left( \bar{x}^{o_{k:k'-1}}_{k'|k'-1,a_{k:k'-1}}, P^{o_{k:k'-1}}_{k'|k'-1,a_{k:k'-1}} \right) \\ \times r^{o_{k:k'-1}}_{k'|k'-1,a_{k:k'-1}} & o_{k'}= 1 \\ \\
        1 - r^{o_{k:k'-1}}_{k'|k'-1,a_{k:k'-1}} \\ \times \overline{p}^D_{a_k'}\left( \bar{x}^{o_{k:k'-1}}_{k'|k'-1,a_{k:k'-1}}, P^{o_{k:k'-1}}_{k'|k'-1,a_{k:k'-1}} \right) & o_{k'} = 0\\
    \end{cases}
\end{multline}
where the first subscript term of the probability of existence $r$, mean $\bar{x}$ and covariance $P$ indicates whether it is predicted $k'|k'-1$ or updated $k'|k'$, the second subscript term denotes the sequence of actions and the superscript includes the sequence of detections/misdetections.
The expected MSGOSPA cost at time-step $k'$ for actions $a_{k:k'}$ and observations $o_{k:k'}$ is
    \begin{multline}\label{eqn: expected MSGOSPA cost}
    C(a_{k:k'}, o_{k:k'}) = \\
    \begin{cases}\!
         \frac{c^2}{2}r^{o_{k:k'}}_{k'|k', a_{k:k'}} & r^{o_{k:k'}}_{k'|k', a_{k:k'}} \leq \Gamma^*_d\\
        \begin{aligned}[c]
                  & \frac{c^2}{2}(1 - r^{o_{k:k'}}_{k'|k', a_{k:k'}}) \\  & +r^{o_{k:k'}}_{k'|k', a_{k:k'}} \min (\text{tr}(P^{o_{k:k'}}_{k'|k', a_{k:k'}}), c^2)
        \end{aligned} 
        & r^{o_{k:k'}}_{k'|k', a_{k:k'}} > \Gamma^*_d
    \end{cases}
\end{multline}
Then, the Bellman optimisation in \eqref{eqn: J nested part 2}, given the assumptions stated at the beginning of Section \ref{sec: non-myopic}, can be simplified as indicated in \eqref{eqn: J nested new notation}. We can see that the internal MSGOSPA integrals in \eqref{eqn: J nested part 2} have been substituted by the bound in Lemma \ref{lemma 1}, (explicitly given by \eqref{eqn: expected MSGOSPA cost}). In addition, the integrals w.r.t. to the measurements in \eqref{eqn: J nested part 2} are now sums, as we are working under the assumption of only having two possible measurement values, and their probability values are given by \eqref{eqn: observing a measurement density}.
\begin{table*}
\centering
\begin{minipage}{0.9\textwidth}
\hrule
\begin{multline}\label{eqn: J nested new notation}
    \hat{J}_{\mu^*_{k:K}(\cdot)} =
    \min_{a_k}\sum_{o_k \in \{0,1\}}  [p(o_{k}|a_{k})C(a_{k},o_{k})
     + \lambda \min_{a_{k+1}} \sum_{o_{k+1} \in \{0,1\}}[ p(o_{k+1}|a_{k:k+1}, o_k)C(a_{k:k+1},o_{k:k+1})\\
     + \ldots + \lambda^{K-k+1}\min_{a_K} \sum_{o_{K} \in \{0,1\}}[ p(o_{K}|a_{k:K-1}, o_{k:K-1})
    C(a_{k:K},o_{k:K}) \bigg] \ldots \bigg] 
\end{multline}
\medskip
\hrule
\end{minipage}
\end{table*}
It should be noted that the predicted mean is independent of the value of $o_k$ and the sequence of actions. Therefore, for the mean, we just need to make the (possibly-multi step) prediction $x_{k'|k}$.
\section{Monte Carlo Tree Search Implementation}\label{sec: MCTS}
Even with the assumptions stated in Section \ref{sec: non-myopic}, the resulting optimisation problem for non-myopic planning \eqref{eqn: J nested new notation} is computationally complex for large time horizons or large action spaces. In this section, we propose the use of MCTS to obtain an approximation to the optimal action \cite{BostroemRost2021}.

MCTS is a selective search algorithm. It incrementally builds a search tree and decides where to explore next based on what it believes to be the most promising avenues. The tree is continually grown until some predefined terminating condition is reached (such as number of iterations). Once this has been reached, the tree returns the best child of the root node, having considered a larger number of actions within the planning horizon. Typically, MCTS is used to maximise the expected reward over an action space, rather than in this case where we minimise an expected cost. However, we can define the cost as the negative of the reward. We proceed to explain the tree structure, the MCTS algorithm and provide an illustrative example on how it works.
\subsection{Tree Structure}\label{sec: tree structure}
The search tree is initialised by creating the root node, which represents the current time-step $k$, and setting its reward $\mathrm{R} = 0$ (or cost $-\mathrm{R} = 0$) and visit count $n = 0$. In \cite{Silver2010}, the tree is viewed as having two different types of nodes, one set corresponding to actions and the other set corresponding to observations. This is due to the MCTS algorithm being applied to a POMDP. The optimal policy is conditioned on both actions and observations, but we present a simplification of this approach with computational advantages, similar to that in \cite{Salvagnini2015}. The simplification we propose and implement is that each node in the tree only represents an action taken at a specific time-step. Each of these action nodes contains the aggregated costs for both receiving a measurement and not (considering and aggregating both observation hypotheses into a single cost) and information on the target state conditioned on past actions. At each update step required in the search tree, we combine the updated Bernoulli density with no measurement and with measurement into a single Bernoulli with a Gaussian single-target density via moment matching, which corresponds to a KL divergence minimisation \cite{Fontana2023}.

Let $r_{k'|k', a_{k:k'}}, \bar{x}_{k'|k', a_{k:k'}}$ and $P_{k'|k', a_{k:k'}}$ be the probability of existence, mean and covariance of each node. How these quantities are predicted, updated and merged (after the update with detection and misdetection) is explained in Appendix \ref{appendix D}. Then, in this setting, each node in the tree contains this information 
\begin{itemize}
    \item Visit count $n$.
    \item Expected cost of visiting node $-\mathrm{R}$, defined in  \eqref{eqn: total cost} (not updated by simulation outcomes).
    \item Expected cost of visiting node across all time-steps $-\mathrm{\bar{R}}$ (updated by simulation outcomes using \eqref{eqn: update node cost}).
    \item Parent node.
    \item Set of child nodes $J$.
    \item Set of available actions $\mathbb{A}_k$.
    \item Target mean $\bar{x}_{k'|k',a_{k:k'}}$.
    \item Target covariance matrix $P_{k'|k',a_{k:k'}}$.
    \item Probability of existence $r_{k'|k', a_{k:k'}}$.
    \item Probability of detection \\$\overline{p}^D_{a_k'}( \bar{x}_{k'|k'-1,a_{k:k'-1}}, P_{k'|k'-1,a_{k:k'-1}})$.
    \item Node depth in global tree.
\end{itemize}
The cost of visiting each node $-\mathrm{R}$ is the cost considering both hypotheses (receiving a target generated measurement, or not) and is calculated using \eqref{eqn: observing a measurement density} and \eqref{eqn: expected MSGOSPA cost}
\begin{multline}\label{eqn: total cost}
-\mathrm{R} = \\
\bigg(1-r_{k'|k'-1, a_{k:k'}} \overline{p}^D_{a_{k'}}\bigg( \bar{x}_{k'|k'-1,a_{k:k'-1}}, P_{k'|k'-1,a_{k:k'-1}}\bigg)\bigg)\\
{C}\bigg(r_{k'|k', a_{k:k'}}^{o_{k'} = 0},P_{k'|k',a_{k:k'}}^{o_{k'} = 0}\bigg)\\
+ \overline{p}^D_{a_{k'}}\bigg( \bar{x}_{k'|k'-1,a_{k:k'-1}}, P_{k'|k'-1,a_{k:k'-1}}\bigg) r_{k'|k'-1, a_{k:k'}}\\
{C}\bigg(r_{k'|k', a_{k:k'}}^{o_{k'} = 1},P_{k'|k',a_{k:k'}}^{o_{k'} = 1}\bigg)
\end{multline}
where the factors before $C(\cdot,\cdot)$ represent the probability of no detection and detection for the current node in the tree and the superscript terms indicate whether a measurement was received $o_k = 1$ or not $o_k = 0$.

\subsection{MCTS Algorithm}
There are four stages to the MCTS algorithm, they are Selection, Expansion, Simulation and Back-propagation \cite{Browne2012}.
\subsubsection{Selection}
In this stage, a node that still has unvisited children is selected. This node will then be used in the next stage, expansion. To select this node, this phase begins at the root node of the tree. Existing nodes are then recursively selected using the Upper Confidence Bound for Trees (UCT) \eqref{eqn: UCT} until an existing node is reached, with at least one unvisited child \cite{Browne2012}. A child node $j$ of the current node is selected according to
\begin{equation}\label{eqn: UCT}
    \underset{j \in J}{\text{arg max}} \bigg\{\mathrm{\bar{R}}_j + 2 \epsilon \sqrt{\frac{\ln n}{n_j}} \bigg\}
\end{equation}
where $n_j$ is the visit count of the child node, $n$ is the visit count of the current node, $\epsilon$ is the trade-off parameter between exploration and exploitation, $J$ is the set of children of the current node and $\mathrm{\bar{R}}_j$ is the expected reward from visiting the child node, which will be explained in the backpropagation phase (see \eqref{eqn: update node cost}).
The UCT equation \eqref{eqn: UCT} considers both the reward of visiting the node and also how many times it has been visited before, with a trade-off between exploration and exploitation, balanced by $\epsilon$. The UCT causes the tree to grow in areas which show high reward, whilst ensuring that is does not get stuck in this area of the search.
The average reward $\mathrm{\bar{R}}_j$ is updated during the back-propagation stage. It should be noted that, at the beginning, the selection stage will always select the root node until all of its children have been added to the tree.
\subsubsection{Expansion}
A new child node $j$ is added to the selected node. This consists of randomly selecting a previously untried action, predicting the target density, generating the synthetic measurement set, calculating the reward of selecting this action and then updating the target density (considering both the measurement and no measurement hypotheses). How both of these hypotheses are considered is described in \eqref{eqn: total r}-\eqref{eqn: total covariance} in Appendix \ref{appendix D}. These node statistics are then used to initialise a new node object which is added to the tree.
\subsubsection{Simulation}
A simulation is then run, starting from the child node $j$. This simulation is run in accordance with the roll-out policy, which is often chosen to be a random path of actions taken down the tree until it reaches a pre-specified terminating condition, which we choose to be the planning horizon. The total reward $\Delta$ of this action sequence is calculated as the discounted sum of all of the costs in this action sequence, beginning at the root node of the tree, ending at the final node of the simulation (at the planning horizon). The total reward $\Delta$ is given by \eqref{eqn: J nested new notation}, with the associated approximation in calculating the updated Bernoulli density (see Subsection \ref{sec: tree structure}) for the sequence of actions $a_{k:K}$ that this path in the tree represents. By starting at the root node, there is always a fair comparison of rewards as the sequence of actions is always the same length.

It can be noted here that the nodes that are used to create a path to the planning horizon, from the current child node $j$ are not added to the tree, and therefore they do not have any node characteristics to update in the back-propagation phase of the MCTS.
\subsubsection{Back-propagation}
The visit count $n$ and associated reward $\bar{R}$ of each of the parent nodes that have been used are updated. The visit count is incremented by one for all parent nodes on the path to the current (child) node $j$. The reward is updated in each node using \eqref{eqn: update node cost}, in which each term refers to the same node that is being updated
\begin{equation}\label{eqn: update node cost}
    \mathrm{\bar{R}}_{new} = \frac{(\mathrm{\bar{R}}_{old} \cdot n) +\Delta}{n + 1}
\end{equation}
where $\mathrm{\bar{R}}_{new}$ is the updated reward associated with this action/node, $\mathrm{\bar{R}_{old}}$ is the reward associated with the action/node prior to this back-propagation phase, $n$ is the visit count of the node and $\Delta$ is the reward calculated from the simulation phase. It can be noted here that the reward is updated prior to incrementing the visit count.
\subsection{Illustrative Example}
As previously discussed, there are four stages to the MCTS algorithm. We will provide an illustrative example of these four stages. At each time-step, consider a sensor with five available actions at each time-step, and therefore a maximum of five child nodes per parent node. The root node represents the time-step $k$ and contains all of the information listed in Section \ref{sec: MCTS}.
\subsubsection{Selection}
\begin{figure}[h]
    \centering
    \includegraphics[scale = 0.25]{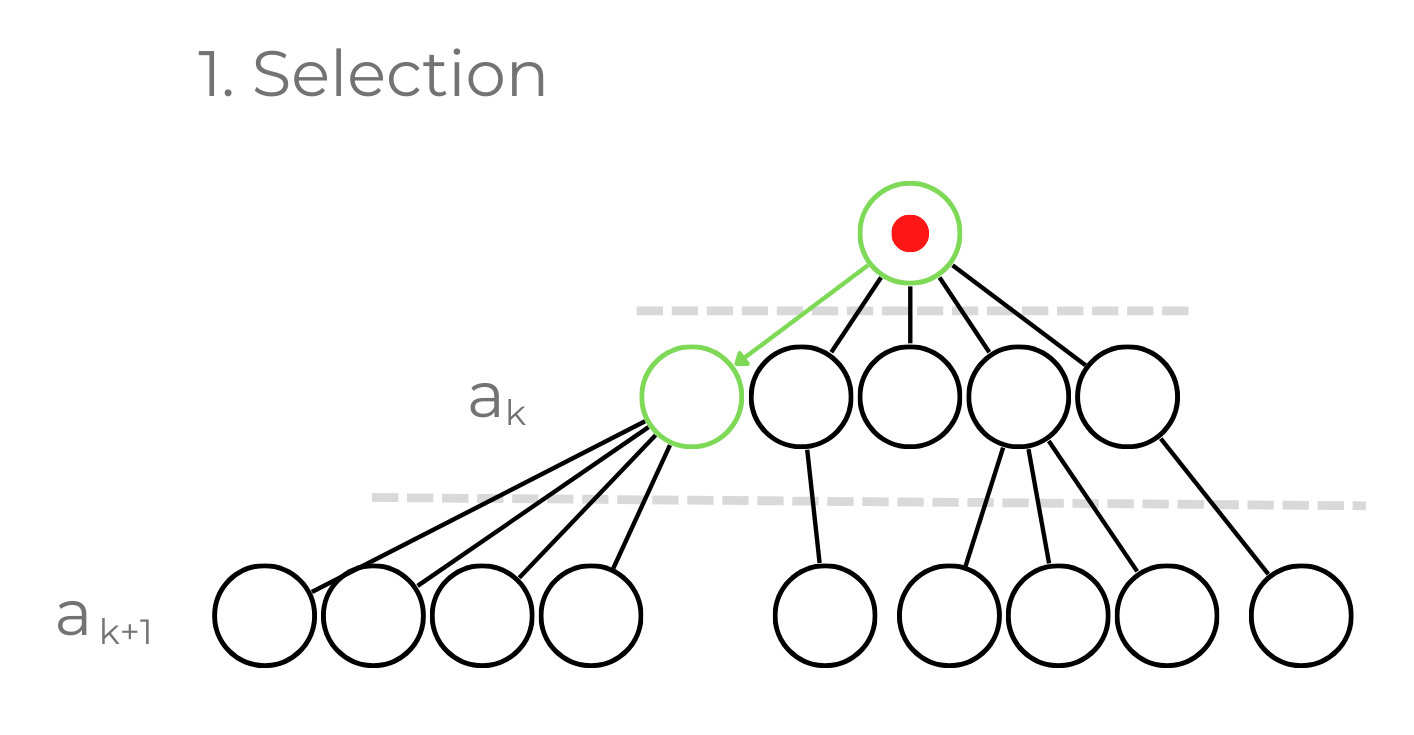}
    \caption{Selection stage of the MCTS. Top node representing the current time-step with the green nodes indicating which have been selected from the pre-existing tree (black nodes).}
    \label{fig: MCTS selection}
\end{figure}

Here, we start at the root node and select a child node based on the UCT criteria set out by \eqref{eqn: UCT}. As can be seen in Figure \ref{fig: MCTS selection}, the node on the farthest left has been selected. In this example, as the first selected node has unexplored children, we do not need to evaluate the UCT again during this MCTS iteration. 
\subsubsection{Expansion}
\begin{figure}[h]
    \centering
    \includegraphics[scale=0.25]{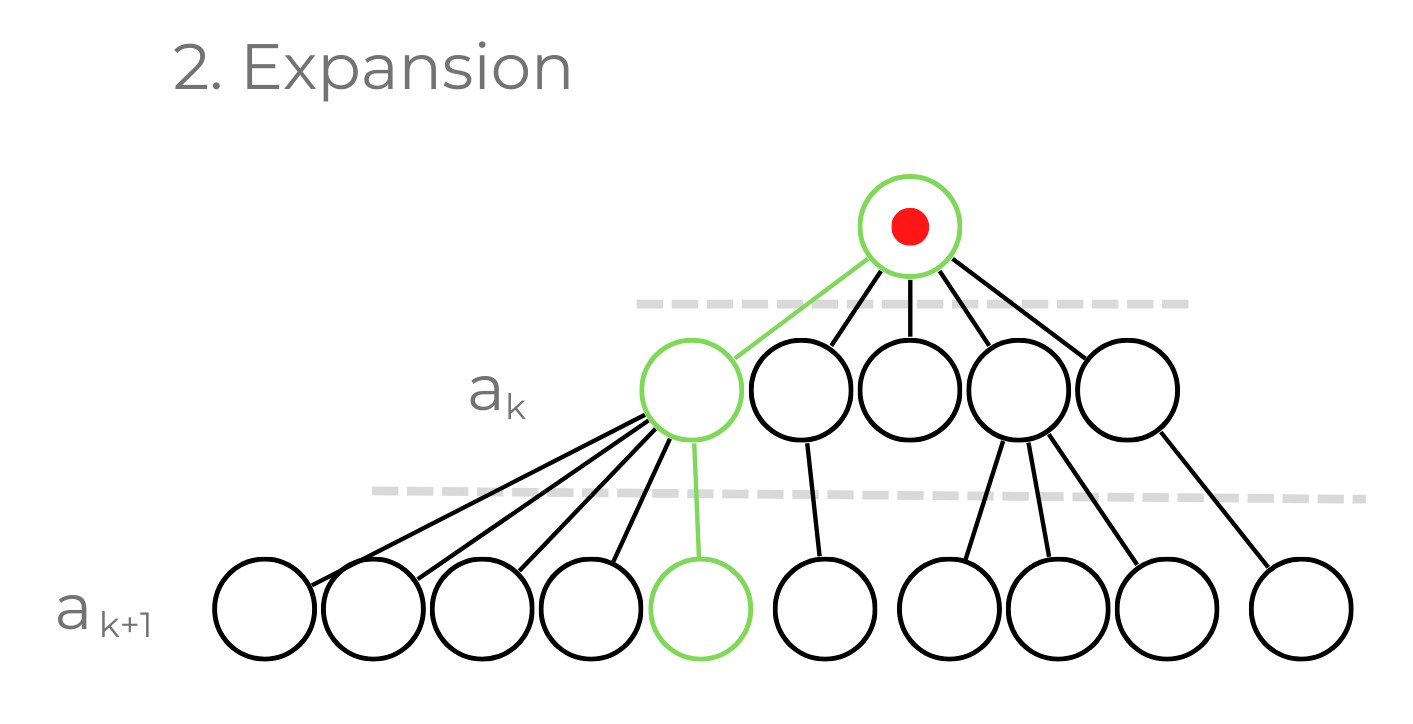}
    \caption{Expansion stage of the MCTS with green nodes indicating those that have been used to get to the point where the tree has been expanded (lowest green node).}
    \label{fig: MCTS expansion}
\end{figure}

Now the parent node has been selected, we add the new child node (representing a sensor action) to the tree in Figure \ref{fig: MCTS expansion}. In this case, the parent node only has one child (action) left to be added/considered and therefore no choice of which child node is required. However, if there were multiple child nodes still to be expanded from the selected node, one would be chosen at random for the expansion. During this phase, the target density is predicted, the ideal measurement set is generated, the cost associated with taking this action (considering both a detection and misdetection hypothesis) is calculated and the target density is then updated (also considering both hypotheses). The node is then added to the tree, initialising its visit count to $0$ and cost $-\mathrm{R}$ calculated as the expected MSGOSPA \eqref{eqn: total cost}.
\subsubsection{Simulation}
Starting at the most recently added node, a simulation is run down to the planning horizon. The route taken by the simulation is governed by the rollout policy, which is chosen to be a random selection of actions down the search tree until the planning horizon is reached. 

\begin{figure}[h]
    \centering
    \includegraphics[scale = 0.25]{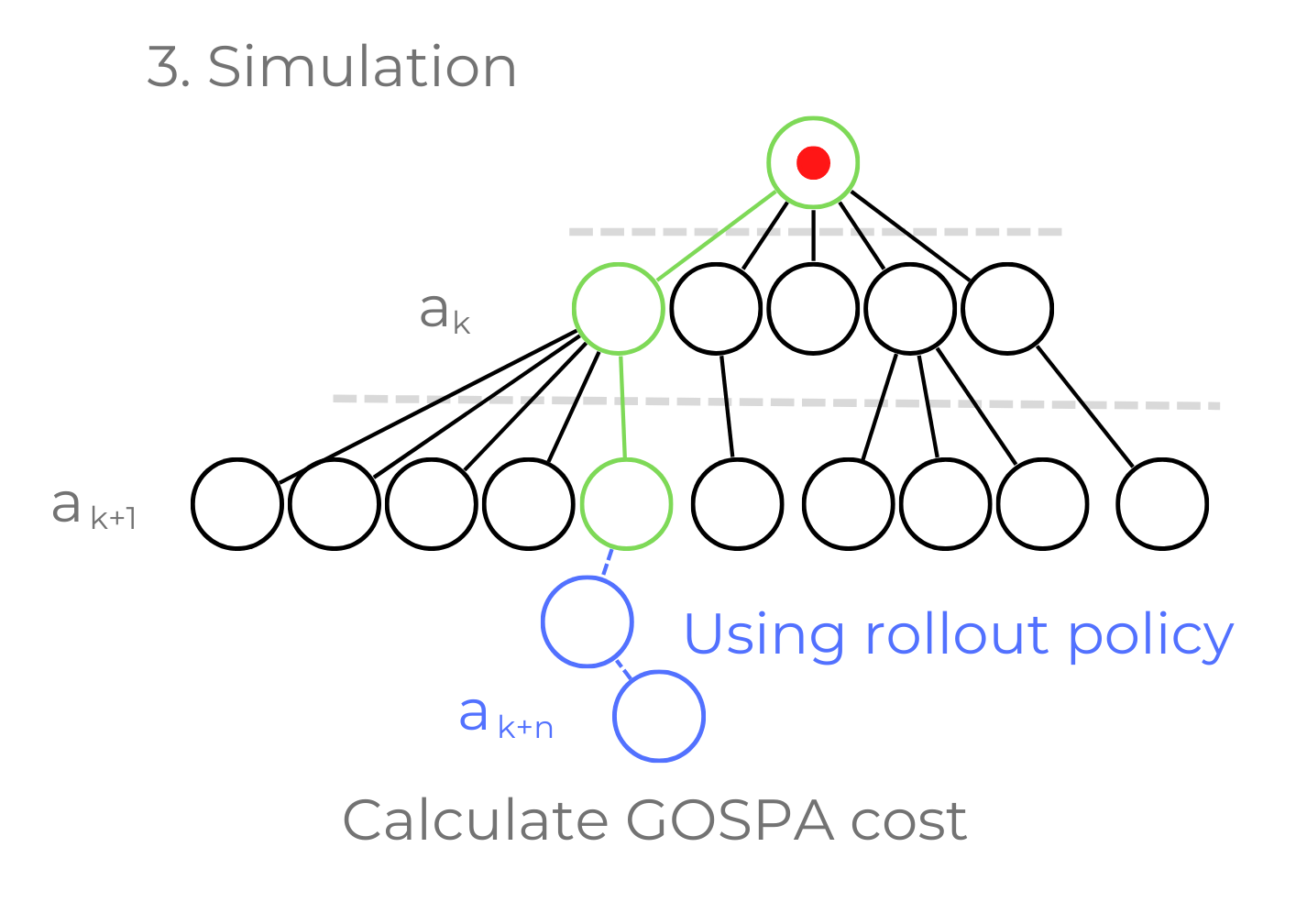}
    \caption{Simulation phase of the MCTS where the green nodes indicate which nodes have been selected (or expanded - lowest green node) and the blue nodes indicating a random path, as governed by the rollout policy, down to the planning horizon.}
    \label{fig: MCTS simulation}
\end{figure}

This phase is illustrated in Figure \ref{fig: MCTS simulation}, it shows an example planning horizon depth of four which determines the maximum search depth of the simulation. As we expanded a node onto the second depth layer of the tree, the simulation phase would only have to look ahead two time-steps to reach the terminating depth. Once the simulation phase has terminated, the cost of this action sequence $\Delta$ is taken to be the discounted sum of all actions, starting at the root node, following the path laid out by the selection and simulation phase, down to the planning horizon.
The nodes depicted in blue in Figure \ref{fig: MCTS simulation} are not added to the tree and therefore their node characteristics are not updated in the next phase as they are not part of the tree, only the expanded node (lowest green node) has been added to the tree.
Note, it is the costs that have not been updated by the other simulation outcomes ($-\mathrm{R}$) that are used in the summation to calculate $\Delta$, and not $-\mathrm{\bar{R}}$.
\subsubsection{Back-propagation}
\begin{figure}[h]
    \centering
    \includegraphics[scale = 0.25]{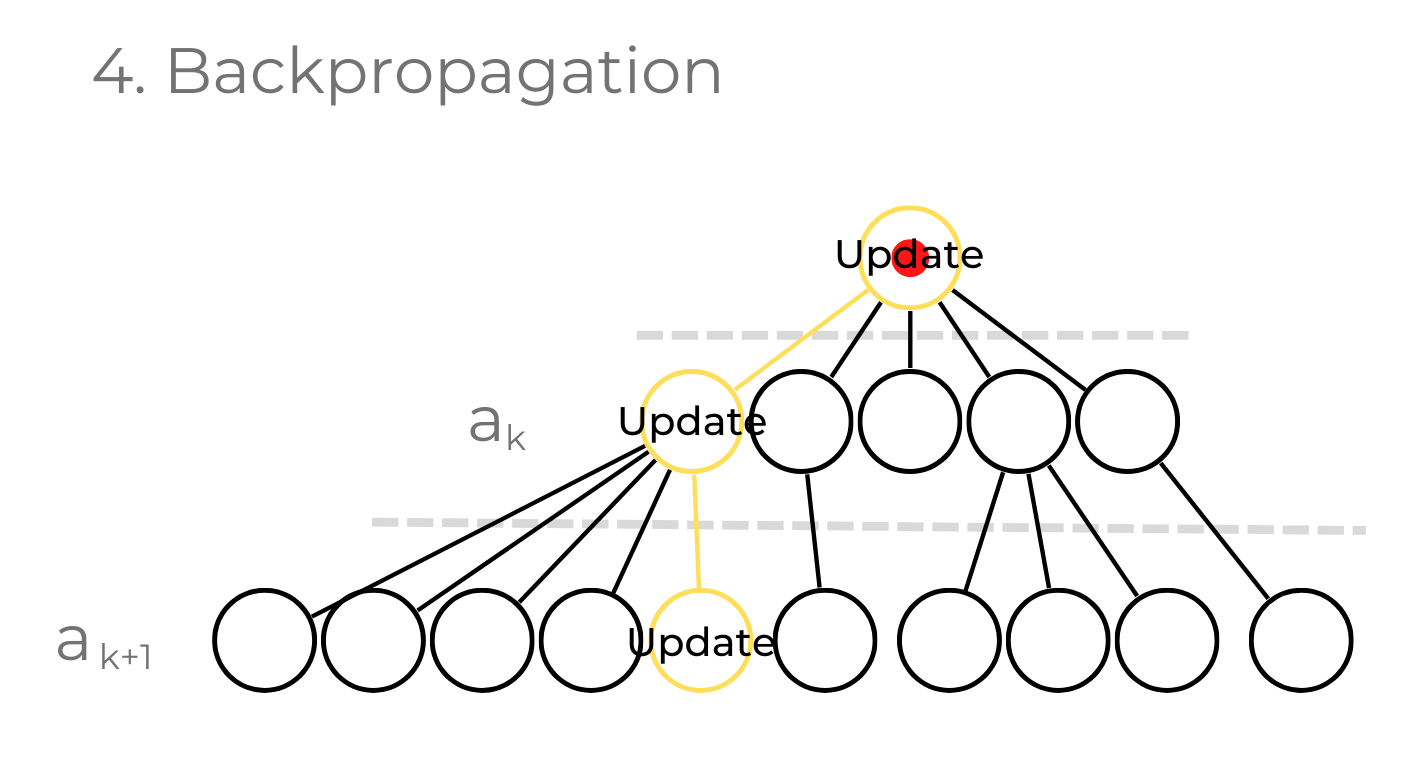}
    \caption{Back-propagation phase of the MCTS with the yellow nodes indicating which node statistics are set to be updated, stating from the lowest yellow node and propagating back up the tree.}
    \label{fig: MCTS backpropagation}
\end{figure}

Here, the node statistics are updated. The final cost computed in the simulation phase $\Delta$ is absorbed into all of the yellow highlighted nodes in Figure \ref{fig: MCTS backpropagation}, in accordance with \eqref{eqn: update node cost}. The visit count of all the yellow highlighted nodes is then incremented by one.
\subsubsection{MCTS Example Summary}
Phases 1 - 4 that have been outlined above constitute a single iteration of the MCTS algorithm. The number of MCTS iterations is limited by either a maximum run time  or computational constraints. In our case, we have limited the computational budget by defining how large the tree can grow. The higher the budget, the more computation required for each time-step as the tree is growing larger each time and looking in more areas of the search space; meaning further into the future.
\section{Numerical Experiments}\label{sec: numerical experiments}
In this section, numerical experiments comparing the proposed algorithm for sensor management with a heuristic policy and an information theoretic algorithm are shown. We outline the sensor movement model, obstacles in the surveillance area, the probability of detection approximation and the tracking performance results from the simulations.

In the non-myopic case, the discount factor $\lambda = 0.7$ was applied to the future rewards \eqref{eqn: J nested new notation}. The chosen rollout policy for the MCTS was a random selection of actions down to a maximum search depth of 10 time-steps. The trade-off parameter was $\epsilon = 0.05$, favouring exploitation over exploration.

\subsection{Target Motion Model}
The target motion is modeled using the nearly constant velocity model \cite[Chap. 6]{BarShalom2004}. The state vector is $x = [p_x, v_x, p_y, v_y]^T$ where $x_p = [p_x, p_y]^T$ is the position vector and $x_v = [v_x, v_y]^T$ is the velocity vector. The transition matrix and process noise covariance matrix are
\begin{equation}\label{eqn: transition and process noise matrices}
F = 
\begin{bmatrix}
    1 & \tau & 0 & 0\\
    0 & 1 & 0 & 0\\
    0 & 0 & 1 & \tau\\
    0 & 0 & 0 & 1\\
\end{bmatrix}
 ; 
Q = q \cdot
 \begin{bmatrix}
    \frac{\tau^3}{3} & \frac{\tau^2}{2} & 0 & 0\\
    \frac{\tau^2}{2} & \tau & 0 & 0\\
    0 & 0 & \frac{\tau^3}{3} & \frac{\tau^2}{2}\\
    0 & 0 & \frac{\tau^2}{2} & \tau\\
\end{bmatrix}
\end{equation}
where $\tau = 1$ and $q = 5$. The birth mean vector is $\bar{x}_B = [0.1, 0, 0.1, 0]^T$ and the birth covariance matrix is diagonal and given by $P_B = \text{diag}([1000, 100, 1000, 100]^T)$.
\subsection{Sensor Model}\label{sec: Sensor model}
\subsubsection{Sensor Movement Model}\label{sec: Sensor movement model}
\begin{figure}[h]
    \centering
    \includegraphics[scale=0.15]{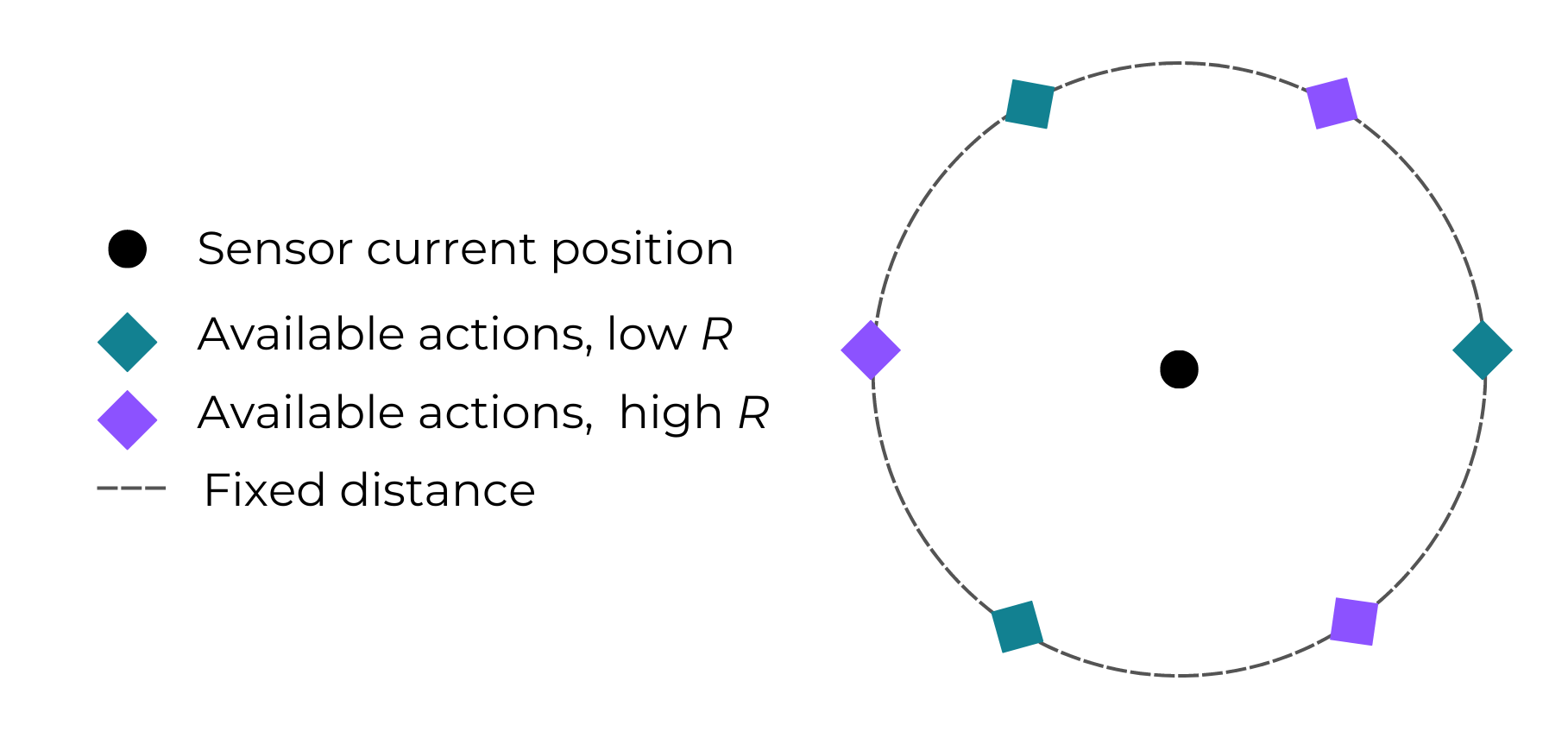}
    \caption{The sensor movement model where the number of available actions is equal to six.}
    \label{fig: sensor movement model}
\end{figure}

We consider a scenario in which the sensor has a limited FOV that can move to keep the target in the FOV. The governing dynamics of the sensor to maintain track of the target is as follows. The sensor has a fixed number of available actions, all equidistant from its current position. Figure \ref{fig: sensor movement model} shows a sensor with $6$ actions. Each action has an associated change in the measurement noise matrix $R$.

This means that the actions available to the sensor can have an impact on the measurement dynamics, as some actions lead to more uncertainty in the measurement (high $R$) than others (low $R$). The sensor has a FOV which is smaller than the area of surveillance region, it is a circular FOV, centred around the sensors current position. The actions are evenly distributed around the circumference of a circle defined by the radius (step size) of the sensor, meaning that the sensor is modelled under constant speed dynamics, capable of travelling a set distance within each discrete time-step $k$. Figure \ref{fig: sensor movement model} depicts the sensor movement model if the number of available actions is equal to $|\mathbb{A}| = 6$.
\subsubsection{Sensor Field of View}\label{sec: sensor field of view}
Let us consider that the sensor FOV is a circle such that the probability of detection is given by
\begin{align}\label{eqn: sensor FOV}
p_{a_{k}}^{D}\left(x\right) & =\begin{cases}
p^{D} & \left\Vert x_{p}-s_{a_{k}}\right\Vert \leq\delta\\
0 & \left\Vert x_{p}-s_{a_{k}}\right\Vert >\delta
\end{cases}
\end{align}
where $\delta$ is the radius and $s_{a_k}$ is the centre of the FOV, i.e. the sensor position.

The linear measurement model consists of the observation matrix $H$ and observation covariance matrix $R$. Considering that we measure target position, the $H$ and $R$ matrices are
\begin{equation}
\label{eqn: observation and observation noise matrices}
H = 
\begin{bmatrix}
    1 & 0 & 0 & 0\\
    0 & 0 & 1 & 0\\
\end{bmatrix}
 ; 
R = 
 \begin{bmatrix}
    10 & 0\\
    0 & 10\\
\end{bmatrix}
\end{equation}
Clutter within the FOV is drawn from a Poisson distribution with a clutter rate $\lambda_c = 1$.
Depending on the action selected, the entries in the $R$ matrix vary. We define a `low $R$'  and a `high $R$' as having entries of 10 and 50 along the main diagonal respectively. An example of actions with different $R$ matrices is shown in Figure \ref{fig: sensor movement model}.
\subsection{Obstacles in the Surveillance Area}
The surveillance area is defined as square, with the sensor being able to move in accordance to the model defined in Section \ref{sec: Sensor movement model}. Once the target leaves the surveillance area, the probability of existence is set to zero and the target is considered to have died. Within this surveillance area, we have also introduced some obstacles to make the sensor management problem more challenging. These obstacles have been designed to impact the sensors ability to move, but not the target. An example of this would be if the sensor platform was a ground vehicle and the target an aerial one, encountering different obstacles in their terrain. If the target enters these regions, the movement dynamics are not affected and neither are the measurements.

\subsection{Probability of Detection Approximation}
A standard approach to approximate the expected probability of detection $\overline{p}^D_{a_k}(\bar{x},P)$ in \eqref{eqn: predicted density of the measurement} in Gaussian filtering is by using its value at the mean $\bar{x}$. In this section, we present an approximation with higher accuracy based on importance sampling that improves the sensor management algorithm. Using \eqref{eqn: sensor FOV}, we can calculate the expected value in \eqref{eqn: expected probability of detection} such that
\begin{align}\label{eqn: pd approximation}
\bar{p}_{a_{k}}^{D} & = p^{D}\int_{\left\Vert x_{p}-s_{a_{k}}\right\Vert \leq\delta}\mathcal{N}\left(x_{p};\overline{x}_{p,k|k-1},P_{p,k|k-1}\right)dx_{p}
\end{align}
where $\overline{x}_{p,k|k-1}$ and $P_{p,k|k-1}$ are the predicted mean and covariance matrix of the positional elements and we have dropped the dependence on $\bar{p}_{a_{k}}^{D}$ on the predicted mean and covariance for notational simplicity.

It is possible to approximate \eqref{eqn: pd approximation} by drawing samples from the Gaussian density. However, the approximation will be inaccurate if the mass of the Gaussian density is far from the FOV. Therefore, we use importance sampling to improve the accuracy of the estimation. In particular, we write \eqref{eqn: pd approximation} as
\begin{multline}
\bar{p}_{a_{k}}^{D} =  \\
  p^{D}\int\mathcal{N}\left(x_{p};\overline{x}_{p,k|k-1},P_{p,k|k-1}\right)\chi_{\left\Vert x_{p}-s_{a_{k}}\right\Vert \leq\delta}\left(x_{p}\right)dx_{p}
\end{multline}
where $\chi_{A}\left(\cdot\right)$ is the indicator function on the
set $A$. To make a uniform density (so that we can draw samples),
we need to account for a normalisation factor. In $n$ dimensions, the volume of a $n$-sphere is
\begin{align}
V_{n}\left(\delta\right) & =\frac{\pi^{n/2}}{\Gamma\left(n/2+1\right)}\delta^{n}
\end{align}
where $\Gamma(\cdot)$ is the Gamma function. In the standard case in which the position is 2-dimensional, we have
\begin{align}
V_{2}\left(\delta\right) & =\pi\delta^{2}
\end{align}
which is the surface of the (circular) FOV in \eqref{eqn: sensor FOV}. Therefore, the uniform density in the field of view is 
\begin{align}
u_{s_{a_{k}},\delta}\left(x_{p}\right) & =\frac{1}{V_{n}\left(\delta\right)}\chi_{\left\Vert x_{p}-s_{a_{k}}\right\Vert \leq\delta}\left(x_{p}\right)
\end{align}
Then, we have
\begin{align}
\bar{p}_{a_{k}}^{D} & = p^{D}V_{2}\left(\delta\right)\int\mathcal{N}\left(x_{p};\overline{x}_{p,k|k-1},P_{p,k|k-1}\right)u_{s_{a_{k}},\delta}\left(x_{p}\right)dx_{p}\\
 & \approx p^{D}V_{2}\left(\delta\right)\frac{1}{I}\sum_{i=1}^{I}\mathcal{N}\left(x_{p,i};\overline{x}_{p,k|k-1},P_{p,k|k-1}\right)
\end{align}
where $x_{p,i}$ is the $i$-th sample from  $u_{s_{a_{k}},\delta}\left(x_{p}\right)$
and $I$ is the number of samples. As $I$ increases, the accuracy of the approximation increases. Specifically, the error of the approximation is $O(I^{-\frac{1}{2}})$ \cite{liu_book2008}.

\begin{figure}[h]
    \centering
    \includegraphics[scale = 0.35]{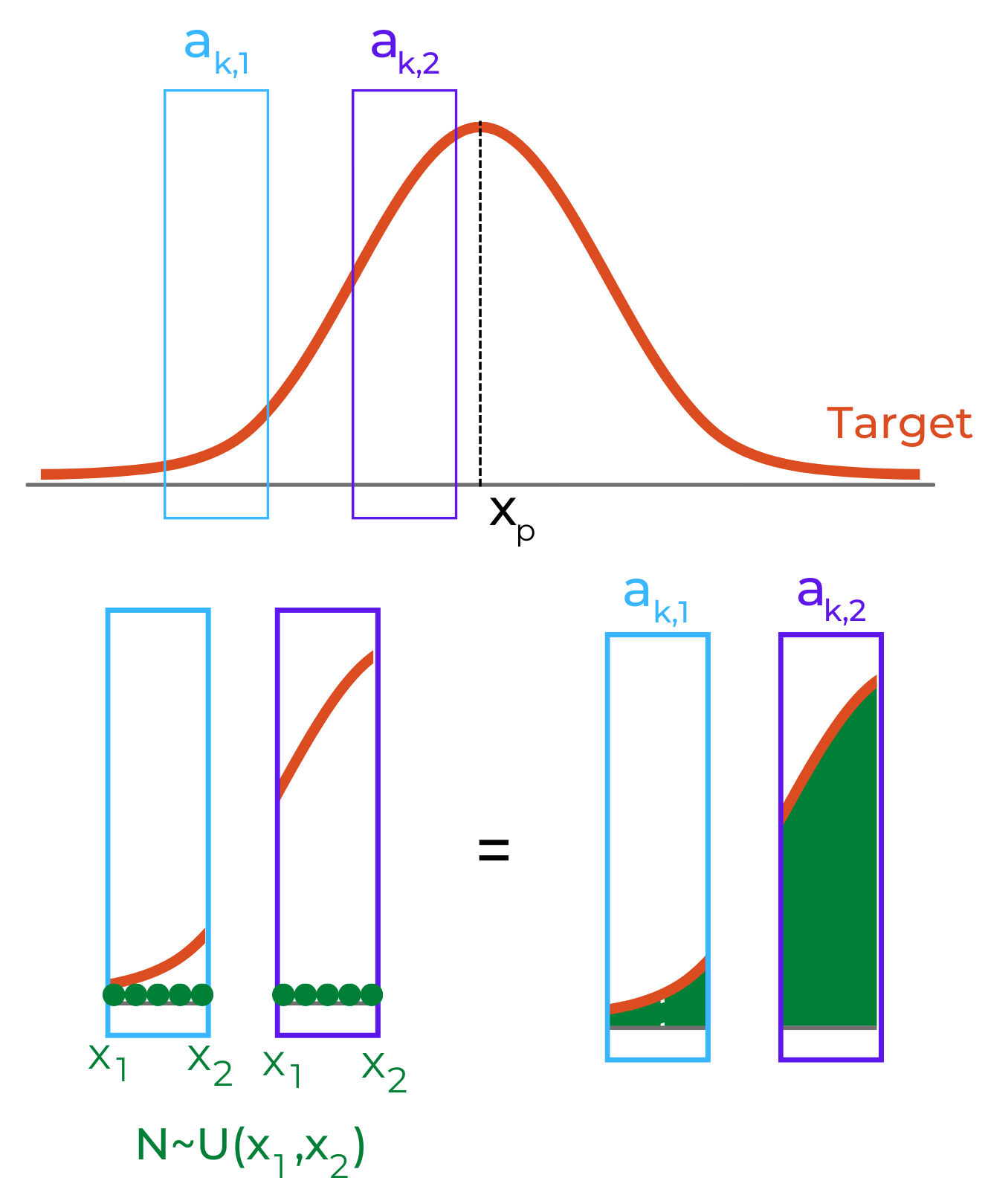}
    \caption{The methodology to calculate the expected probability of detection, depicted in 1D using 5 samples. The top section denoting the target distribution and two action windows ($a_{k,1}$ and $a_{k,2}$), representing the FOV of two actions. The bottom left figure shows the 5 uniform samples being drawn between the upper and lower limits of each action window. The bottom right figure illustrates the samples being evaluated through the target distribution and an average of them taken.}
    \label{fig: expected prob detection graphic}
\end{figure}

The expected probability of detection of the target is calculated as a function of both the target and potential sensor positions (actions) at the current time-step $k$. Figure \ref{fig: expected prob detection graphic} illustrates the methodology in which sampling is used to calculate an expected probability of detection in one dimension.

Each action ($a_{k}$) can be visualised as a window that can see some part of the underlying target distribution. We can calculate an expected probability of detection by calculating the average value of the target distribution which is encapsulated within each window (action). We propose to do this by generating a fixed amount of uniformly distributed samples within each window, evaluate each of these w.r.t. the target distribution, and then take an average of their values. This way, we are able to have a separate probability of detection for each action, based on where the sensor would be relative to the target distribution.
\subsection{Results}
In this section, we evaluate the performance of the following algorithms in two scenarios, one in which there are obstacles in the surveillance region, and one where there are none. The first algorithm is the nearest sensor, sensor management algorithm (NS) which can be considered as the heuristic solution to the basic problem where there are no obstacles. The NS algorithm selects the mean of the highest weighted Gaussian component of the predicted density and moves the sensor to the available action that is closest in Euclidean distance to it. This algorithm cannot be extended to non-myopic planning as it does not provide a cost or reward for each action. The second is the myopic GOSPA-driven  algorithm, where the performance metric GOSPA is used as a driver for myopic sensor management. The third one is a (myopic) information theoretic sensor management algorithm, using the highest weighted component in the prediction, see Appendix \ref{Appendix C}. Finally, we show three variations of the non-myopic GOSPA-driven algorithm implemented via MCTS, one with a smaller computational budget of 10 nodes, one with a budget of 50 and one with a higher budget of 150 nodes per tree. Note, the smallest budget MCTS has been omitted from the Figures \ref{fig:GOSPA Plots no obstacles} and \ref{fig:GOSPA Plots obstacles} as to not clutter them.

The root mean square GOSPA (RMS-GOSPA) error at time-step $k$ is 
\begin{equation}\label{eqn: RMS-GOSPA}
\sqrt{\mathrm{E}\left[d^{2}\left(X_{k},\hat{X}_{k}\right)\right]}	\eqsim\sqrt{\frac{1}{n}\sum_{i=1}^{n}d^{2}\left(X_{k},\hat{X}_{k,i}\right)}
\end{equation}
where $\hat{X}_{k,i}$ is the estimated set of targets at time-step $k$ in the $i$-th Monte Carlo (MC) run, and $n$ is the number of Monte Carlo runs. 
As a measure of performance, we also calculate the RMS-GOSPA error across all time-steps defined as
\begin{equation}
\sqrt{\frac{1}{K}\sum_{k=1}^{K}\mathrm{E}\left[d^{2}\left(X_{k},\hat{X}_{k}\right)\right]}	\eqsim\sqrt{\frac{1}{Kn}\sum_{k=1}^{K}\sum_{i=1}^{n}d^{2}\left(X_{k},\hat{X}_{k,i}\right)}
\end{equation}
where $K$ is the number of time-steps in the simulation.

The RMS-GOSPA errors were calculated  over a simulation length of 300 time-steps and 80 Monte Carlo runs. The GOSPA parameter $c=2r$ where $r=40$ and is the radius of the sensors FOV. Two scenarios (obstacles and no obstacles) are simulated with different ground truth trajectories. The planning horizon is set to $5$ and therefore the action space has a maximum of $6^5$ options to search through.
\subsubsection{No Obstacles}
\begin{figure}
    \centering
    \includegraphics[width=\linewidth]{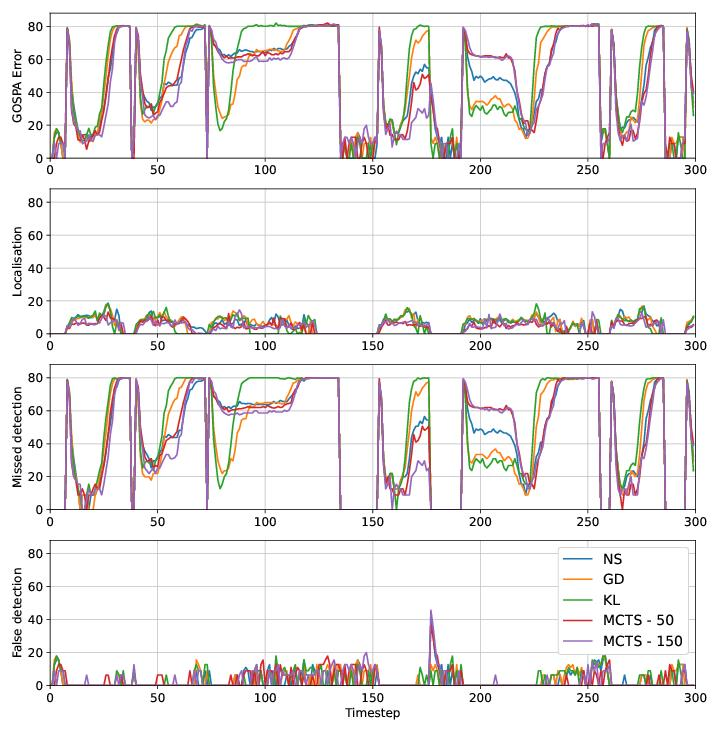}
    \caption{RMS-GOSPA error breakdown for each time-step where no obstacles are in the surveillance area. Labeled a - d starting from the top plot (a) GOSPA error, (b) localisation error contributions, (c) missed detection error contributions, (d) false detection error contributions}
    \label{fig:GOSPA Plots no obstacles}
\end{figure}

The simulated surveillance region for the no-obstacle simulation is shown in Figure \ref{fig: scenario_snapshot_no_obstacles}. The GOSPA error plots are shown in Figure \ref{fig:GOSPA Plots no obstacles} for the NS algorithm, myopic GOSPA-driven (GD), the information theoretic KL (KL) and MCTS non-myopic GOSPA-driven (MCTS-50, MCTS-150) algorithms.
As can be seen from the GOSPA error plots in Figure \ref{fig:GOSPA Plots no obstacles}, the MCTS - 150 is the most performant. However, the performance is not largely greater than the rest of the tested algorithms which provide a computationally less demanding solution. In Table \ref{tab: results no obstacles}, we show the RMS-GOSPA errors across all time steps, also including different values of $\lambda$ for MCTS-10.
\begin{figure}[h]
    \centering
    \includegraphics[scale = 0.5]{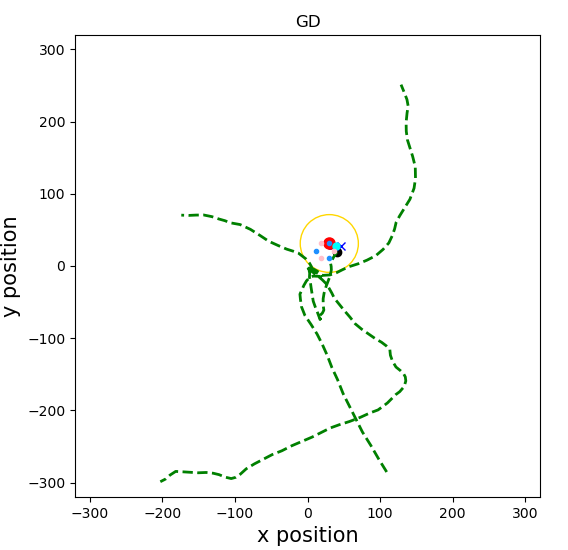}
    \caption{Scenario snapshot of the GD algorithm in the final MC run when there are no obstacles. The red circle indicates the sensors current position surrounded by the yellow circle which is the sensors current FOV. The pink circles indicate an available action with low $R$ and the blue high $R$.  The black dot is the targets current state, the green lines are legacy tracks where the target has been. The pale blue circle shows the extracted target state from the Bernoulli filter and the dark blue cross shows the measurements received at this time-step. Here there are no major benefits to using a more computationally demanding non-myopic approach.}
    \label{fig: scenario snapshot no obstacles}
\end{figure}
\begin{table}[h]
\caption{RMS-GOSPA error for each algorithm, with no obstacles} 
\label{tab: results no obstacles}
\begin{center}
\begin{tabular}{c|c}
\hline
\textbf{Algorithm} & \textbf{Avg. GOSPA Error} \\ \hline 
    NS                 &          45.53    \\ 
     KL                 &          48.15    \\
    GD   (MCTS $\lambda=0$)   &       45.52  \\ 
    
    MCTS - 10 ($\lambda=0.7$)       &          44.46    \\ 
     MCTS - 50 ($\lambda=0.7$)         &          44.47    \\ 
    MCTS - 150 ($\lambda=0.7$)  & 44.43 \\ 
    MCTS - 10 ($\lambda=0.5$)  &  44.39 \\ 
    MCTS - 10 ($\lambda=0.1$)  &  42.90 \\ \hline   
\end{tabular}
\end{center}
\end{table}
From these simulations, it can be seen that in such a simple scenario, there is no benefit to utilising non-myopic approaches as they are more computationally demanding and are similarly performant compared to simpler algorithms.
\subsubsection{Obstacles}
\begin{figure}[h]
    \centering
    \includegraphics[width=\linewidth]{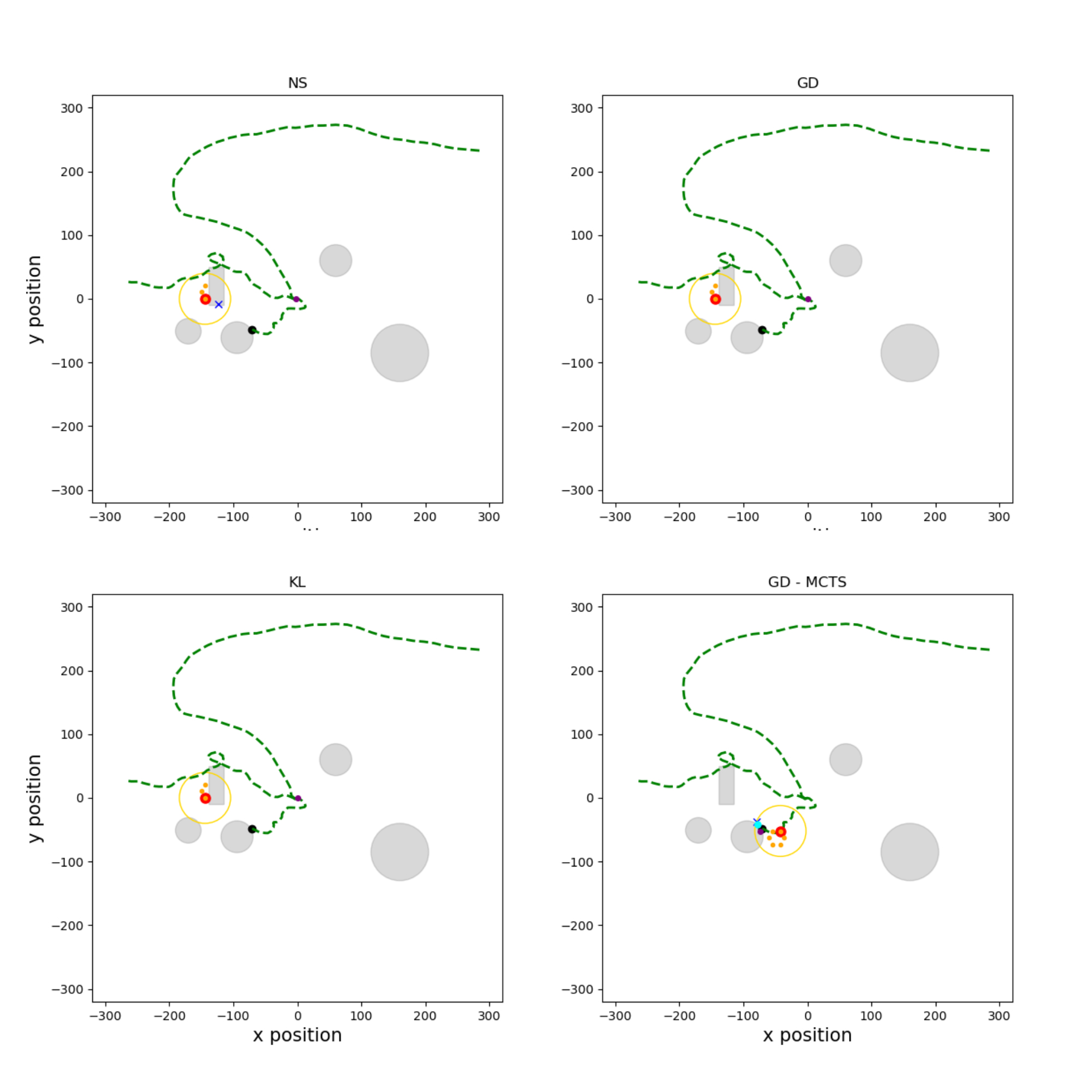}
    \caption{Snapshot of the final MC run for all algorithms at $k=210$ where obstacles for the sensor (not the target) are present in the surveillance area (shown as grey shapes). Top left - Nearest Sensor, Top right - GOSPA Driven, Bottom left - Kullback-Leibler Driven, Bottom right - GOSPA Driven MCTS (non-myopic). Here, the obstacles are visualised by the grey shapes. The green dashed line represents the target track and the black dot the target. The red marker is the sensors current location with the yellow circle surrounding it representing the sensors FOV. The light blue dot is the state estimate from the filter and the blue crosses represent measurements (both clutter and target generated). In myopic approaches, the sensor gets stuck in an obstacle and loses the target. The non-myopic GOSPA-driven approach can navigate around the obstacle and keep the target in the FOV.}
    \label{fig: scenario snapshot obstacles}
\end{figure}

\begin{figure}[h]
    \centering
    \includegraphics[width = \linewidth]{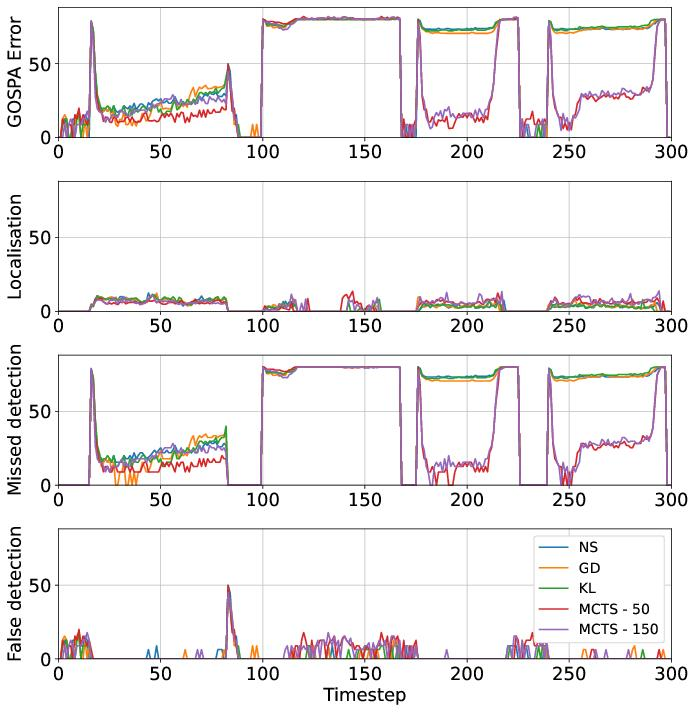}
    \caption{RMS-GOSPA error breakdown for each time-step where obstacles are in the surveillance area. Labeled a - d starting from the top plot (a) GOSPA error, (b) localisation error contributions, (c) missed detection error contributions, (d) false detection error contributions.}
    \label{fig:GOSPA Plots obstacles}
\end{figure}

Figure \ref{fig: scenario snapshot obstacles}, a snapshot of the simulation at time step $k=210$, illustrates a case where myopic approaches were unable to navigate around an obstacle (shown in grey) and therefore got stuck behind the obstacle when navigating back to the birth location. In contrast, the non-myopic algorithm(s) were able to navigate around the obstacles. The myopic approaches were found to be unable to maintain track of any targets after the point that they got stuck behind an obstacle. This usually happened at around time-step $k=95$, hence after this point, all of the myopic approaches remain at almost maximum GOSPA error, as shown in Figure \ref{fig:GOSPA Plots obstacles}. 

The GOSPA cost broken down into its constituent parts is shown in Figure \ref{fig:GOSPA Plots obstacles}. Peaks in the missed detection error value were usually due to the target  appearing in the surveillance area, but are yet to be detected by the sensor. 
For each simulation the average GOSPA error has been collated in Table \ref{tab: results obstacles}. The approximate average error of the myopic approaches is around $50$, whilst the non-myopic approach is around $35$. As would be expected, the larger the planning horizon for the non-myopic approaches (i.e. the further into the future the tree could look and grow), the better the tracking performance. We also found that the larger the obstacles were relative to the sensors movement step size, the larger the budget required to enable the non-myopic approach to find a path that navigated around them. Within the simulated scenario, there is no discernible difference in performance between the three non-myopic budgets 10, 50 and 150). This is because for each simulation, the obstacles and sensor movement speed are unchanged. As the budget required to navigate around an obstacle is proportional to the obstacle size and the sensor step size, even a modest budget of 10 is able to plan around the obstacles in the given simulation - resulting in a similar performance of all three differing budgets where $\lambda=0.7$. For $\lambda=0.7$ and $\lambda=0.5$, the algorithm has similar performance. Where $\lambda=0.1$, the performance of the algorithm drops as it is not sufficiently weighting future costs to be able to navigate around the obstacle.
\begin{table}[h]
\caption{RMS-GOSPA error for each algorithm, with obstacles} 
\label{tab: results obstacles}
\begin{center}
\begin{tabular}{c|c}
\hline
\textbf{Algorithm} & \textbf{Avg. GOSPA Error} \\ \hline 
    NS                 &   51.67     \\ 
    KL       &   51.14               \\  
    GD (MCTS $\lambda=0$)      &   50.19 \\  
    
    MCTS - 10 ($\lambda=0.7$)      & 35.81                 \\
     MCTS - 50   ($\lambda=0.7$)       &  35.52        \\ 
    MCTS - 150 ($\lambda=0.7$) & 35.01 \\ 
    MCTS - 10 ($\lambda=0.5$)  &   34.31\\ 
    MCTS - 10 ($\lambda=0.1$)  &  48.79 \\ \hline         
\end{tabular}
\end{center}
\end{table}

Table \ref{tab: computation time} shows the wall clock time of one 300 time-step simulation and also the time per time-step. The simulations were run on an Intel(R) core (TM) i5-10210U CPU @ 1.6GHz 2.11GHz and the code was written in object-oriented Python. As expected, myopic approaches have shorter run times, and we can control the computational complexity of MCTS via its computational budget.

\begin{table}[h]
\caption{Wall clock computation time for each algorithm, with obstacles} 
\label{tab: computation time}
\begin{center}
\begin{tabular}{c|c|c}
\hline
\textbf{Algorithm} & \textbf{Avg. time [s]} &\textbf{Avg. time per time-step [s]}\\ \hline 
    NS                 &   13.6        &    0.05   \\ 
    KL                 &   36.24         &   0.12   \\ 
    GD                 &   40.3         &    0.13  \\  
    MCTS - 10         &    64.0       &         0.21 \\
     MCTS - 50          &  220.4          & 0.73 \\ 
    MCTS - 150      &  561.9 &  1.87  \\ \hline         
\end{tabular}
\end{center}
\end{table}

\section{Conclusion}\label{sec: conclusion}
In this paper, we have proposed a framework for both myopic and non-myopic GOSPA driven sensor management for Bernoulli filtering, a suitable approximation for implementing the planning algorithm based on Gaussian single-target distributions and an upper bound on the MSGOSPA error, and the development of an MCTS method to efficiently conduct non-myopic GOSPA driven sensor management.

We have shown that in a Bernoulli scenario where there are no obstacles, the GD algorithm performs similarly to the heuristic solution to the problem (NS). We have also shown that the metric driven approach is more performant than an information theoretic approach (KL). We have also demonstrated, and provided a detailed summary of, how non-myopic sensor management can be conducted using an MCTS with GOSPA as a driver, and provided a scenario in which there is a clear benefit to planning non-myopically.

In future work, we plan to extend this work to include multiple targets, using Poisson multi-Bernoulli mixture filters \cite{BostroemRost2021, BostroemRost2022}. Another line of future research is to develop reinforcement learning algorithms in combination with Bayesian multi-target tracking algorithms. Another area to explore in further work is to implement parallelised versions of the MCTS algorithm to improve computation time, such as those described in \cite{Chaslot2008}.

\begin{table*}[h]
\centering
\begin{minipage}{0.9\textwidth}
\hrule
\begin{multline}\label{eqn: appendix r}
        r_{k|k,a_k}^1  = \frac{1 - \bar{p}^D_{a_k}(\bar{x}_{k|k-1}, P_{k|k-1})(1 - \frac{\mathcal{N}(z;, \hat{z}_{a_k},S_{a_k})}{\lambda_c(z)})}{1 - r_{k|k-1}\bar{p}^D_{a_k}(\bar{x}_{k|k-1}, P_{k|k-1})(1 - \frac{\mathcal{N}(z;, \hat{z}_{a_k},S_{a_k})}{\lambda_c(z)})}r_{k|k-1} \\
         = \frac{\lambda_c(z) - \bar{p}^D_{a_k}(\bar{x}_{k|k-1}, P_{k|k-1})(\lambda_c(z) - \mathcal{N}(z;, \hat{z}_{a_k},S_{a_k}))}{\lambda_c(z) - r_{k|k-1}\bar{p}^D_{a_k}(\bar{x}_{k|k-1}, P_{k|k-1})(\lambda_c(z) - \mathcal{N}(z;, \hat{z}_{a_k},S_{a_k}))} 
\end{multline}
\medskip
\hrule
\end{minipage}
\end{table*}

\bibliographystyle{abbrv}
\bibliography{References}

%
%



%

\appendices

\section{Set Integral}\label{appendix A}
Let $f\left(\cdot\right)$ be a real-valued function on the space $\mathcal{F}\left(\mathbb{R}^{n_{x}}\right)$, which denotes the set of all the finite subsets of $\mathbb{R}^{n_{x}}$. Then, its set integral is defined as \cite{Mahler_book14}
\begin{equation}
    \int f(X)\delta X=\sum_{n=0}^{\infty}\frac{1}{n!}\int\ldots\int f\left(\left\{ x_{1},...,x_{n}\right\} \right)dx_{1}\ldots dx_{n}
\end{equation}

We can see that the set integral sums over all possible set cardinalities,
represented by the integer $n$. Then, for cardinality $n$, the set
integral performs $n$ integrals over the space $\mathbb{R}^{n_{x}}$.
More details on the set integral are provided for example in \cite{Mahler_book14}.


\section{Proof of Lemma 1}\label{Appendix B}
In this Appendix, we provide the proof of Lemma 1. The MSGOSPA error is
\begin{multline}\label{eqn: MSGOSPA error}
    \int d^{2}\left(X_{k},\hat{X}\left(a_{k},Z_{k}\right)\right)f_{k|k}\left(X_{k}\left|Z_{k};a_{k}\right.\right) \delta X_{k}\\
     = d^{2}\left(\emptyset,\hat{X}\left(a_{k},Z_{k}\right)\right)\left(1-r_{k|k,a_{k}}^{|Z_{k}|}\right)\\
     \quad+r_{k|k,a_{k}}^{|Z_{k}|}\int d^{2}\left(\left\{ x_{k}\right\} ,\hat{X}\left(a_{k},Z_{k}\right)\right)p_{k|k}\left(x_{k}\right)dx_{k}
\end{multline}
If $r_{k|k,a_{k}}^{|Z_{k}|}\leq\Gamma_{d}$, we substitute \eqref{eqn: estimator} into \eqref{eqn: MSGOSPA error} to obtain 
\begin{equation}
    \int d^{2}\left(X_{k},\emptyset\right)f_{k|k}\left(X_{k}\left|Z_{k};a_{k}\right.\right)\delta X_{k}
    =r_{k|k,a_{k}}^{|Z_{k}|}\frac{c^{2}}{2}
\end{equation}
which proves the first entry of \eqref{eqn: decomposed cost}. 

If $r_{k|k,a_{k}}^{|Z_{k}|}\geq\Gamma_{d}$, we substitute \eqref{eqn: estimator} into \eqref{eqn: MSGOSPA error} to obtain
\begin{equation}
\begin{split} & \int d^{2}\left(X_{k},\hat{X}\left(a_{k},Z_{k}\right)\right)f_{k|k}\left(X_{k}\left|Z_{k};a_{k}\right.\right)\delta X_{k}\\
 & =\frac{c^{2}}{2}\left(1-r_{k|k,a_{k}}^{|Z_{k}|}\right)+r_{k|k,a_{k}}^{|Z_{k}|}\\
 & \times\int\min\left(c^{2},\left\Vert x_{k}-\overline{x}_{k|k,a_{k}}^{|Z_{k}|}\right\Vert ^{2}\right)\mathcal{N}\left(x_{k};\overline{x}_{k|k,a_{k}}^{|Z_{k}|},P_{k|k,a_{k}}^{|Z_{k}|}\right)dx_{k}
\end{split}
\label{eq:bound_derivation1}
\end{equation}

Now, we can obtain two upper bounds to the integral in the above expression.
Since $\min\left(c^{2},\left\Vert x_{k}-\overline{x}_{k|k,a_{k}}^{|Z_{k}|}\right\Vert ^{2}\right)\leq c^{2}$,
we first have that
\begin{align}
 & \int\min\left(c^{2},\left\Vert x_{k}-\overline{x}_{k|k,a_{k}}^{|Z_{k}|}\right\Vert ^{2}\right)\mathcal{N}\left(x_{k};\overline{x}_{k|k,a_{k}}^{|Z_{k}|},P_{k|k,a_{k}}^{|Z_{k}|}\right)dx_{k}\nonumber \\
 & \leq c^{2}\label{eq:inequality_append1}
\end{align}
On the other hand, we have
\begin{alignat}{1}
\min\left(c^{2},\left\Vert x_{k}-\overline{x}_{k|k,a_{k}}^{|Z_{k}|}\right\Vert ^{2}\right) & \leq\left\Vert x_{k}-\overline{x}_{k|k,a_{k}}^{|Z_{k}|}\right\Vert ^{2}
\end{alignat}
This inequality implies that
\begin{align}
 & \int\min\left(c^{2},\left\Vert x_{k}-\overline{x}_{k|k,a_{k}}^{|Z_{k}|}\right\Vert ^{2}\right)\mathcal{N}\left(x_{k};\overline{x}_{k|k,a_{k}}^{|Z_{k}|},P_{k|k,a_{k}}^{|Z_{k}|}\right)dx_{k}\nonumber \\
 & \leq\int\left\Vert x_{k}-\overline{x}_{k|k,a_{k}}^{|Z_{k}|}\right\Vert ^{2}\mathcal{N}\left(x_{k};\overline{x}_{k|k,a_{k}}^{|Z_{k}|},P_{k|k,a_{k}}^{|Z_{k}|}\right)dx_{k}\nonumber \\
 & =\mathrm{tr}\left(P_{k|k,a_{k}}^{|Z_{k}|}\right)\label{eq:inequality_append2}
\end{align}

Therefore, we can write the inequalities (\ref{eq:inequality_append1})
and (\ref{eq:inequality_append2}) compactly as
\begin{align}
 & \int\min\left(c^{2},\left\Vert x_{k}-\overline{x}_{k|k,a_{k}}^{|Z_{k}|}\right\Vert ^{2}\right)\mathcal{N}\left(x_{k};\overline{x}_{k|k,a_{k}}^{|Z_{k}|},P_{k|k,a_{k}}^{|Z_{k}|}\right)dx_{k}\nonumber \\
 & \leq\min\left(c^{2},\mathrm{tr}\left(P_{k|k,a_{k}}^{|Z_{k}|}\right)\right)
\end{align}
Substituting this expression into (\ref{eq:bound_derivation1}), we
obtain
\begin{align}
 & \int d^{2}\left(X_{k},\hat{X}\left(a_{k},Z_{k}\right)\right)f_{k|k}\left(X_{k}\left|Z_{k};a_{k}\right.\right)\delta X_{k}\nonumber \\
 & \leq\frac{c^{2}}{2}\left(1-r_{k|k,a_{k}}^{|Z_{k}|}\right)+r_{k|k,a_{k}}^{|Z_{k}|}\min\left(c^{2},\mathrm{tr}\left(P_{k|k,a_{k}}^{|Z_{k}|}\right)\right)
\end{align}
which proves the second entry in \eqref{eqn: decomposed cost} and finishes the proof of Lemma \ref{lemma 1}.

\section{Kullback-Leibler Divergence between two Bernoulli Gaussian Densities}\label{Appendix C}
For completeness, here we provide the equation used to calculate the KL divergence between two Bernoulli Gaussian densities. In particular, in information-theoretic sensor management, we calculate the KL divergence between the posterior density and the predicted density \cite{Kreucher07}. Therefore, we can use the following expressions instead of the upper bound on the MSGOSPA cost in Lemma 1 to determine sensor actions.

Let ${f}_{k|k'}(\cdot)$ with $k'\in \{k, k-1\}$ be the posterior and predicted Bernoulli densities with probability of existence $r_{k|k'}$ and Gaussian single-target density with mean $\bar{x}_{k|k'}$ and covariance matrix $P_{k|k'}$. If the probability of existence $r_{k|k-1} \notin \{0,1\}$, the KL divergence of ${f}_{k|k}$ from ${f}_{k|k-1}$ is given by \cite{Fontana2023}
\begin{multline}
    D_{KL}({f}_{k|k}||{f}_{k|k-1})\\
    = (1-r_{k|k-1})\log \frac{1 - r_{k|k-1}}{1 - r_{k|k}} + r_{k|k-1}\log\frac{r_{k|k-1}}{r_{k|k}}\\
    + \frac{r_{k|k-1}}{2} \bigg[ \text{tr}\bigg((P_{k|k})^{-1}P_{k|k-1}\bigg) - \log \bigg(\frac{|P_{k|k-1}|}{|P_{k|k}|}\bigg) - n_x\\
    + (\bar{x}_{k|k} - \bar{x}_{k|k-1})^T(P_{k|k})^{-1}(\bar{x}_{k|k} - \bar{x}_{k|k-1})\bigg]
\end{multline}
If $r_{k|k-1} = r_{k|k} \in \{0,1\}$ the KL divergence is
\begin{multline}
    D_{KL}\bigg(f_{k|k}||f_{k|k-1}\bigg)\\
     = \frac{r_{k|k-1}}{2}\bigg[ \text{tr} \bigg( (P_{k|k})^{-1} P_{k|k-1} \bigg) - \log\bigg(\frac{|P_{k|k-1}|}{|P_{k|k}|}\bigg) - n_x \\
    + (\bar{x}_{k|k} - \bar{x}_{k|k-1})^T (P_{k|k})^{-1} (\bar{x}_{k|k} - \bar{x}_{k|k-1}) \bigg]
\end{multline}
\section{Probability of Existence, Mean and Covariance in the MCTS implementation}\label{appendix D}
This appendix explains how the target probability of existence, mean and covariance are propagated in the MCTS implementation explained in Section \ref{sec: MCTS}. The probability of existence, mean and covariance of the target are predicted using \eqref{eqn: predicted density}. Then for the update, we update them with no measurement using \eqref{eqn: x^0}-\eqref{eqn: r_k|k^0} giving rise to $r_{k'|k', a_{k:k'}}^{o_k=0}$, $\bar{x}_{k'|k',a_{k:k'}}^{o_k=0}$ and $P_{k'|k',a_{k:k'}}^{o_k=0}$.

The update with measurements is carried out with \eqref{eqn: yes measurement prob existence}-\eqref{eqn: P^1} giving rise to $r_{k'|k', a_{k:k'}}^{o_k=1}$, $\bar{x}_{k'|k',a_{k:k'}}^{o_k=1}$ and $P_{k'|k',a_{k:k'}}^{o_k=1}$.

The two updates (with and without measurements) are merged via KL divergence minimisation as in \cite{Fontana2023}.
\begin{multline}\label{eqn: total r}
    r_{k'|k', a_{k:k'}} =
    w^0 \cdot r_{k'|k', a_{k:k'}}^{o_k=0}
    + w^1 \cdot r_{k'|k', a_{k:k'}}^{o_k=1}
\end{multline}
\begin{multline}\label{eqn: total mean}
    \bar{x}_{k'|k',a_{k:k'}} =
    w^0 \cdot \bar{x}_{k'|k',a_{k:k'}}^{o_k=0}
    + w^1 \cdot \bar{x}_{k'|k',a_{k:k'}}^{o_k=1}
\end{multline}
\begin{multline}\label{eqn: total covariance}
    P_{k'|k',a_{k:k'}} =
    w^0 \cdot P_{k'|k',a_{k:k'}}^{o_k=0}
    + w^1 \cdot P_{k'|k',a_{k:k'}}^{o_k=1}
\end{multline}
where 
\begin{multline}
    w^0 = 1-\bigg(r_{k'|k'-1, a_{k:k'}} \\ \cdot \overline{p}^D_{a_k'}\bigg( \bar{x}_{k'|k'-1,a_{k:k'-1}}, P_{k'|k'-1,a_{k:k'-1}}\bigg)\bigg)
\end{multline}
\begin{multline}
    w^1 = r_{k'|k'-1, a_{k:k'}} \\ \cdot \overline{p}^D_{a_k'}\bigg( \bar{x}_{k'|k'-1,a_{k:k'-1}}, P_{k'|k'-1,a_{k:k'-1}}\bigg)
\end{multline}
and $r_{k'|k', a_{k:k'}}^{o_k=0}$ is the updated probability of existence in a misdetection hypothesis, $r_{k'|k', a_{k:k'}}^{o_k=1}$ is the updated probability of existence in a detection hypothesis. $\bar{x}_{k'|k',a_{k:k'}}^{o_k=0}$ and $\bar{x}_{k'|k',a_{k:k'}}^{o_k=1}$ are the updated target means for the misdetection and detection hypotheses respectively. $P_{k'|k',a_{k:k'}}^{o_k=0}$ and $P_{k'|k',a_{k:k'}}^{o_k=1}$ are the updated target covariance matrices for the misdetection and detection hypothesis, respectively.

\end{document}